\documentclass[aps,prl,twocolumn,superscriptaddress,longbibliography]{revtex4-2}

\usepackage{amsmath}
\usepackage{amssymb}
\usepackage{amsthm}
\usepackage{tikz}
\usepackage{mathrsfs}
\usepackage{graphicx}
\usepackage{color}
\usepackage{xcolor}
\usepackage{hyperref}
\usepackage{cleveref}
\usepackage{physics}
\usepackage{soul}
\usepackage{bm}

\newcommand{\nn}{\nonumber\\}                    	
\newcommand{\dagg}{^{\dagger}}      	            
\newcommand{\ihip}{\text{I}}					

\newcommand{\tS}{\text{S}} 
\newcommand{\tE}{\text{E}} 
 
\newcommand{\trE}{\text{tr}_{\text{E}}} 
\newcommand{\jsd}{{\rm JSD}}

\newcommand{\om}{\omega}

\newcommand{\ep}{\epsilon}

\newcommand{\hgam}{\hat{\gamma}}

\newcommand{\hH}{\hat{H}}

\newcommand{\hsig}{\hat{\sigma}}

\newcommand{\jdoi}{\hat{\varrho}_{\tS \tE}}
\newcommand{\sdo}{\hat{\rho}_{\tS}}
\newcommand{\sdoi}{\hat{\varrho}_{\tS}}

\newcommand{\edo}{\hat{\rho}_{\tE}}
\newcommand{\edoi}{\hat{\varrho}_{\tE}}
\newcommand{\sL}{\mathcal{L}_{\tS}}

\newcommand{\fK}{\mathcal{K}^{(2)}}

\newcommand{\hrho}{\hat{\rho}}
\newcommand{\rate}[1]{\Gamma_{\hspace{-0.5pt}#1}}
\newcommand{\corr}[1]{C_{\hspace{-0.5pt}#1}}
\newcommand{\spec}[1]{S_{\hspace{-0.5pt}#1}}

\begin{document}


\title{Open system probes of renormalization group flow}

\author{Andrew Keefe}
\thanks{These authors contributed equally to this work.}
\affiliation{Department of Physics and Applied Physics, University of Massachusetts, Lowell, MA 01854, USA}
\author{Brenden Bowen}
\thanks{These authors contributed equally to this work.}
\affiliation{Department of Physics and Applied Physics, University of Massachusetts, Lowell, MA 01854, USA}
\author{Saptarshi Biswas}
\affiliation{Department of Physics and Astronomy, Northwestern University, Evanston, IL 60208, USA}
\author{\\ Albion Lawrence}
\affiliation{Martin Fisher School of Physics, Brandeis University, Waltham, MA 02453, USA}
\author{Nishant Agarwal}
\email{nishant_agarwal@uml.edu}
\affiliation{Department of Physics and Applied Physics, University of Massachusetts, Lowell, MA 01854, USA}
\author{Archana Kamal}
\email{archana.kamal@northwestern.edu}
\affiliation{Department of Physics and Applied Physics, University of Massachusetts, Lowell, MA 01854, USA}
\affiliation{Department of Physics and Astronomy, Northwestern University, Evanston, IL 60208, USA}

\begin{abstract}
Open system probes can provide an efficient means to characterize quantum many-body systems by employing them as engineered environments. The key idea is to map long-range spatial correlations of the environment onto dynamical correlations in the evolution of a simple quantum probe. Using the example of a qubit coupled to a transverse-field Ising model, we show how the non-Markovian rate or spectral flow can be used to identify stable and unstable fixed points, infer scaling dimensions of relevant fields, and deduce the renormalization group flow induced by deformations around any fixed point.
\end{abstract}

\maketitle


\textit{Introduction.--} 
Motivated by several breakthroughs in quantum information platforms \cite{Fauseweh2024} and emerging applications such as many-body sensing \cite{Montenegro2025}, there has been a surge in interest to leverage non-unitary effects for probing and manipulating many-body correlations and dynamics \cite{Khemani2016,Swingle2019,Fazio2025}. On the one hand, this allows opportunities to harness the rapidly expanding control and measurement capabilities, connectivity, and size of NISQ-era quantum simulators \cite{Barreiro2011,Choi2017,Mi2022,King2022,Kim2023, MiGoogle2024}; on the other, it continues to spur new techniques for characterizing multi-qubit and many-body correlations \cite{Baez2020,Will2025,Lee2026}. These developments provide an attractive alternative to traditional methods based on direct measurements of long-range correlations in complex quantum systems, such as neutron \cite{Zaliznyak2015} or light \cite{ DevereauxandHackl2007} scattering, multi-particle interferometry \cite{Folling2005} and transport measurements \cite{Reulet2003}, or atomic imaging \cite{Altman2004,Bakr2009}, which are experimentally complicated and computationally intensive to predict. In addition, standard methods remain severely limited in their engineering potential especially when deployed in new material platforms and non-equilibrium scenarios \cite{Bando2020}. 

An alternate approach, inspired by quantum reservoir engineering, is to employ the many-body quantum system of interest as an engineered environment, whose correlations can be inferred via its imprints on the decoherence of a simple probe system. This strategy to use quantum probes has been successfully explored for detecting quantum critical points \cite{Quan2006,Cucchietti2007,Haikka2012} and, more recently, has been extended to detect a variety of quantum phase transitions, including dynamical~\cite{Damski2011} and topological phase transitions~\cite{Giorgi2019}, integrability-to-chaos transitions~\cite{Mirkin2021}, and measurements of many-body excitation spectra \cite{Roberts2024}. Recent works have also shown how non-Markovian spectra of a probe qubit can provide a direct measurement of environment spectral density, though such studies have remained confined to few-mode environments such as those encountered in circuit- or cavity-QED settings \cite{keefe2025}. 

In this letter, we show how a quantum probe coupled to a relevant many-body operator can not only witness phase transitions, but even provide quantitative estimates of scale-invariant properties associated with the renormalization group flow of a many-body system near criticality. Specifically, the relationship between spatial and temporal correlations in many-body systems, dictated by symmetries at critical points, manifests as a direct connection between finite correlation lengths and finite correlation timescales of the many-body environment. Since a finite timescale associated with environmental correlations necessarily leads to non-Markovianity in the probe dynamics, we show how the time-dependent rates or frequency-dependent spectral widths allow inferring quantities such as stability of fixed points and scaling dimensions, via standard quantum optical measurements of the probe. 
%
%
\par
\textit{Central qubit coupled to TFIM.--} As an archetypal example of a quantum many-body system undergoing a phase transition, we consider a 1D transverse-field Ising model (`environment' $E$) interacting with a probe qubit (`system' $S$) described by the full Hamiltonian ${\hH_{\tS \tE} = \hH_{\tS} + \hH_{\tE} + \hH_{\text{int}}}$ ($\hbar=1$),
\begin{subequations}
\label{eq:QICHamiltonian}
\begin{align}
\label{eq:QubitFreeHamiltonian}
	\hH_{\tS}
&=
	\frac{
		\omega_{\tS}
	}{
		2
	}
	\hsig^{z}_{\tS}
	\,,
\\
\label{eq:TICHamiltonian}
	\hH_{\tE}
&=
	-
	J
	\sum_{j}
	\hsig^{z}_{j}
	\hsig^{z}_{j+1}
	-
	h
	\sum_{j}
	\hsig^{x}_{j}
	\,,
\\
\label{eq:IntHamiltonian}
	\hH_{\text{int}}
&=
	\frac{g}{\sqrt{L}}
	\hsig^{z}_{\tS}
	\sum_{j}
	\hsig^{x}_{j}
	\,,
\end{align}
\end{subequations}
where the summations are over the $N = L/a$ spins in the Ising chain, with $L$ being the length of the chain and $a$ the lattice spacing, $ \omega_{\tS} $ is the qubit frequency, $ J $ is the spin-spin coupling, $ h $ is the transverse magnetic field strength, and $g$ is the interaction strength. The TFIM in \cref{eq:TICHamiltonian} has a quantum critical point at $ h = J $. Using the standard Jordan-Wigner transformation~\cite{Jordan_1928}, we map the TFIM Hamiltonian to a system of free fermions that can be readily diagonalized in terms of the eigenoperators $ \hgam_{k_{m}} $ and $ \hgam_{k_{m}}\dagg $ obeying the anti-commutation relation $ \{\hgam_{k_{i}} , \hgam_{k_{j}}\dagg \} = L\delta_{i,j}$ (see \cref{app:diagTFIM}). Here, $ \{ k_{m} = 2 \pi m / L \} $ index the wave-vectors associated with the discrete Fourier modes derived assuming periodic boundary condition. Lastly, we normal order the diagonalized Hamiltonian with respect to the environment state, which, for the quadratic Hamiltonian here, amounts to replacing $ \hH_{\tE} \to \mathop{:} \hH_{\tE} \mathop{:} \equiv \hH_{\tE} - \langle \hH_{\tE} \rangle_{\hrho_{\tE}}~$ \footnote{Beyond quadratic order, normal ordering contains additional terms to cancel all bubble diagrams \cite{Bowen:2025kyo}.}.
In this representation, and under the statistical field theory limit with a finite UV cutoff set by the lattice spacing $a$, the environment operator entering the interaction $\hH_{\text{int}}=\hsig_{\tS}^{z} \hat{\mathcal{O}}_{\tE}$, transforms as \footnote{We normal order $ \hat{\mathcal{O}}_{\tE} $ and move a $ \mathop{:} \hgam_{k}\dagg \hgam_{k}  \mathop{:} $ term that leads to spurious thermal bubbles in the second-order environment correlation function to the free Hamiltonian, where it becomes a system-dependent frequency correction to the TFIM spectrum that ultimately vanishes in the thermodynamic limit (see \cref{app:diagTFIM}).
}
\begin{align}
	\hat{\mathcal{O}}_{\tE}
&=
    \frac{
		i 
		g
	}{
		\sqrt{L}
	}
    \int_{-\pi/a}^{\pi/a} 
    \frac{\dd k}{2 \pi}
	\frac{
		r_{k}
	}{
		\ep_{k}
	}
	\left(
		\hgam_{-k}\dagg
		\hgam_{k}\dagg
		-
		\hgam_{k}
		\hgam_{-k}
	\right)
    \,,
\label{eq:EnvironmentOperator}
\end{align}
with $ r_{k} = 2 J \sin(a k) $ and $\ep_{k} = 2\sqrt{J^{2} + h^{2} - 2hJ\cos(a k)} $. We report results in the quantum field theory limit  ($ a \to 0 $) in the main text, since the resulting expressions are more amenable to analytical evaluation, and present results for finite $a$ in \cref{app:StatisticalFieldTheoryLimit}.
%
\par
\textit{Dynamic probe of scaling dimensions.--} Following standard methods of quantum master equation (QME) construction, we derive a Redfield equation describing the dynamics of the probe qubit after tracing out the TFIM environment, \footnote{The absence of a unitary correction (Lamb shift) in the Redfield equation is a direct result of the QND interaction.}
\begin{align}
	\frac{ 
		\dd 
		\sdo
	}{ 
		\dd t 
	}
&=
	-
	i 
	\frac{\om_{\tS}}{2}
	\left[
		\hsig^{z}_{\tS}
		,
		\sdo(t)
	\right]
	+
	\rate{\mu,T}(t) 
	\left[
		\hsig^{z}_{\tS}
		\sdo(t)
		\hsig^{z}_{\tS}
		-
		\sdo(t)
	\right].
\label{eq:Redfieldequation}
\end{align}
The induced dephasing rate on the qubit,
\begin{align}
	\rate{\mu, T}(t)
&=
	\int_{0}^{t}
	\dd t_{1}
	\Re
	\left[
		\corr{\mu,T}(t,t_{1})
	\right]
	\,,
\label{eq:gamma}
\end{align}
is directly determined by the TFIM correlation function
\begin{align}
    \corr{\mu,T}(t,t_{1})
&=
    2
	\trE
	\big\{
		\hat{\mathcal{O}}_{\tE}\dagg(t)
		\hat{\mathcal{O}}_{\tE}(t_{1})
		\edo(T)
	\big\}
	\,\nonumber\\
&\hspace{-0.7cm} =
	\frac{4 g^{2}}{\pi}
    \hspace{-2pt}
	\int_{|\mu|}^{\infty}
    \hspace{-2pt}
	\dd \ep_{k}
	\sqrt{1-\frac{\mu^{2}}{\ep_{k}^{2}}}
	\frac{\cos[2\ep_{k}(t-t_{1}+\frac{i}{2 T})]}{1+\cosh\left[\ep_{k}/T\right]}
	\,,
\label{eq:EnvironmentCorr}
\end{align}

with $\{ \mu, T \}$ denoting the transverse field-dependent gap and temperature of the TFIM, respectively. In writing the second line of \cref{eq:EnvironmentCorr}, we have substituted \cref{eq:EnvironmentOperator} for $\hat{\mathcal{O}}_{\tE}$ and used Wick's theorem to write the resulting correlation function in terms of two-point correlators, as shown in \cref{app:CorrFunction}. While evaluating \cref{eq:EnvironmentCorr} for general $\{ \mu, T \}$ requires numerical methods, analytical expressions can be obtained in various limiting cases, and we show expressions for the cases $\{ \mu, 0 \}$ and $\{ 0, T \}$ in the appendix as well.

Our first key finding is that the probe qubit exhibits Markovian dephasing with a constant (or zero) positive rate at \emph{all} three fixed points of the TFIM, not only at  at the quantum critical point (QCP), $\{\mu,T\} =\{0,0\}$ \cite{Haikka2012}. 

Notably, this manifestation of Markovianity at the fixed points along the zero-temperature axis, corresponds to two very distinct kinds of environments, namely an infinite-range correlated bath (soft mode with zero gap) and an uncorrelated bath (stiff mode with infinite gap). This can be seen most clearly by reformulating the real part of \cref{eq:EnvironmentCorr} along each axis of the ($\mu, T$) phase diagram in terms of a characteristic time scale associated with the TFIM environment, 
\begin{align}
    C^{\rm (axis)} (\tau)= 2g^2\delta(\tau) - \frac{4g^2}{\tau_{B}} F_{\rm axis}\left(\frac{\tau}{\tau_{B}}\right),
    \label{eq:EnvironmentCorr2}
\end{align}
where $C^{\rm (axis)}(\tau) \in \{ \text{Re}[C_{\mu,0}(\tau)], \text{Re}[C_{0,T}(\tau)] \}$ with ${\tau \equiv t-t_1}$. Here $ F_{\rm axis}(x)$ is a non-negative bounded function for ${x \in [0,\infty)}$ that isolates the non-Markovian contribution to the probe dephasing rate (see \cref{app:CorrFunction}). Along the gap ($\mu$) axis, $\tau_B\equiv 1/(2|\mu|)$, while along the temperature axis, $\tau_B\equiv\pi/(8T)$. Evidently, at the QCP, $\tau_{B} \rightarrow \infty$, suppressing the contribution of the non-Markovian tail to the environment spectral density. On the other hand, as ${\tau_{B} \rightarrow 0}$ for $\mu \to \infty$, the non-Markovian tail collapses into a spike with zero correlation time, leading to ${(1/\tau_{B})F_{\mu}({\tau}/{\tau_{B}}) \rightarrow \delta(\tau)/2}$. Similarly, along the temperature-axis at zero-gap, $(1/\tau_{B}) F_{T}({\tau}/{\tau_{B}}) \rightarrow \delta(\tau)/4$ as $T\to \infty$. The two Markovian limits of $\tau_B = \{\infty, 0\}$ constitute a direct manifestation of non-trivial and trivial fixed points corresponding to infinite and zero correlation length, respectively. It is instructive to note that the $\delta$-correlated $C^{\rm (axis)}(\tau)$ at the QCP is a direct consequence of conformal symmetry  and the dimension of the environment operator, ${\rm dim}[\hat{\mathcal{O}}_{\tE}] =1$. This can be seen explicitly by mapping the TFIM at the QCP to a fermionic conformal field theory (CFT) with power-law correlators in real space and time. As shown in \cref{app:CFTCorrFunction}, spatial integration of the Lorentz-invariant CFT correlator leads to $\corr{0,0}(t,t_{1})$ with a $\delta$-correlated real part, in agreement with \cref{eq:EnvironmentCorr2}. 

\begin{figure}[t!]
    \includegraphics[width=\columnwidth]{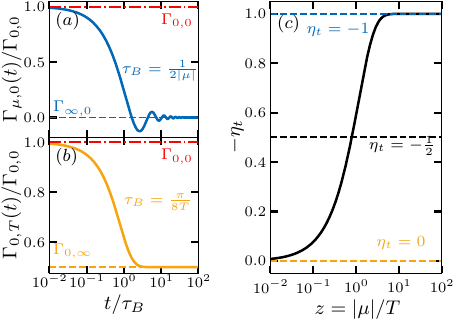}
    \vspace{-20pt}
    \caption{Time-dependent dephasing rates on the probe qubit coupled to the TFIM environment varied along the (a) gap ($\mu$) and (b) temperature ($T$) axes. In both cases, the rate flows from the unstable QCP to the respective stable fixed point along each axis. (c) Exponent determining relative weights of the gap and temperature in the bulk TFIM correlator timescale $\tau_{B}$.
 }
\label{fig:Rates}
\end{figure}

The induced dephasing rate $\rate{\mu, T}(t)$ exhibits explicit time-dependence due to a finite bath timescale $\tau_{B}$ associated with deformations away from the QCP. Along the gap and temperature axes, it can be approximated with an exponential at short time, with the non-Markovian regime set by the respective $\tau_B$, 
\begin{subequations}
\begin{align}
    \label{eq:ShortTimeRatesVac}
	\rate{\mu, 0}(t)
    &\approx
    g^2 e^{-2|\mu|t},
    \\
\label{eq:ShortTimeRatesCrit}
\rate{0, T}(t)
 &\approx
\frac{g^2}{2}\qty(1+e^{-\frac{8 T}{\pi}t})
  \,.
\end{align}
\label{eq:ShortTimeRates}%
\end{subequations}
These expressions show that the probe dephasing rates at the fixed points are determined entirely by the probe-TFIM coupling, consistent with the scale-invariant nature of TFIM correlations at these points. As shown in \cref{fig:Rates}(a)-(b), along the gap axis the dephasing rate flows from $\rate{0, 0} =g^{2}$ to $\rate{\infty,0} = 0$, while along the temperature axis it flows from $\rate{0, 0}=g^{2}$ to $\rate{0,\infty} = g^{2}/2$. It is clear that $\rate{0, 0}$ serves as an upper bound of the dephasing rate induced on the probe qubit as also reported in ref.~\cite{Quan2006}. Further, due to the negative contribution of the non-Markovian tail of the correlator in \cref{eq:EnvironmentCorr2}, the dephasing rate is suppressed with increase in temperature eventually halving to its zero-temperature value at $T \rightarrow \infty$. While this may seem counterintuitive at first, it can be understood from the temperature-dependent weight of the fermion-bilinear correlators which, for translation-invariant Hamiltonians such as the one considered in \cref{eq:IntHamiltonian}, reduce to $n_{k}^{2}(T) + (1- n_{k}(T))^{2}$ where $n_{k}$ is the occupation in the respective Majorana mode (see \cref{app:CorrFunction}). This suppression of rate is analogous to sub-Poissonian fluctuations that manifest in full counting statistics associated with low-temperature mesoscopic transport of non-interacting fermions ~\cite{JongBeenakker}, though here the suppression of fluctuations persists at high temperatures due to nonlinear environment correlations.
%
%
\par
Remarkably, the rates $\rate{\mu, 0}(t)$ and $\rate{0, T}(t)$ mimic the renormalization group flow along $\mu$ and $T$, respectively, with the stable fixed points appearing as the steady state of the non-Markovian rates. This implies that recording the rate flow at a given point in the TFIM phase diagram provides a direct means to infer the relevant scaling dimension associated with a given deformation about the QCP. To this end we generalize the notion of ``rate of rate'', $\tau_{B}^{-1} = -\dot{\Gamma}_{\mu, T}/\Gamma_{\mu, T}$, for an arbitrary deformation in the $\{\mu, T\}$ phase diagram including the crossover region near the critical cone ($\mu = T$). Such a situation may also be relevant for a general many-body environment where short-time correlations are not set by a single time-constant due to multiple competing time-scales. Since non-Markovianity is strongest at short times, $\tau \ll \tau_{B}$ in \cref{eq:EnvironmentCorr2}, we define $\tau_B$  off the TFIM axis via the time constant inferred from the initial acceleration of dephasing at its onset. Using a linear-in-time expansion of $\Gamma_{\mu,T}(t)$, we define an emergent exponent $\eta_t$ that sets the contribution of different timescales obtained from \cref{eq:ShortTimeRates} [\cref{fig:Rates}(c)],
\begin{equation}
    \tau_B=\tau_{\mu}^{-\eta_t}\tau_T^{1+\eta_t},~\eta_t\in[D_{\mu}-2,D_{T}-1],
    \label{eq:mixedtauB}
\end{equation}
with $\tau_{\mu}\equiv (2|\mu|)^{-1}$ and $\tau_T\equiv\pi(8T)^{-1}$. For deformations along the scaling fields (`eigen-axes' of RG flow), $\tau_{B}$ is entirely determined by a single time-constant. Then ${\lambda \sim \tau_{B}^{-D_{\lambda}}}$, with the resultant value of $\eta_{t}$ along the scaling fields, yields the respective scaling dimension $D_{\lambda} = 1$ for $\lambda \in \{\mu, T\}$ (see \cref{app:CorrFunction}). 

It is worthwhile to comment on the validity of the QME approach adopted here, especially near the QCP. At first glance, the notion of weak-coupling underlying the QME construction may seem problematic given the probe coupling to a relevant operator in \cref{eq:IntHamiltonian}. The situation may seem even more pathological in the absence of any finite energy denominator contribution, from either the TFIM environment near the QCP or the probe qubit given the QND nature of the probe-TFIM interaction, needed to regularize the perturbation series. Contrary to these naive expectations, since the environment state and fluctuations remain Gaussian at all times, the QME construction to second-order in coupling ($g$) is sufficient to compute exact rates. We validate this by computing the rate $\rate{\mu, 0}(t)$ in the zero temperature limit using the Loschmidt echo \cite{Quan2006,Haikka2012}, which permits an exact evaluation of decoherence rate as long as the total Hamiltonian can be written in terms of system projectors [see \cref{app:LoschEcho}],
\begin{align}
	\rate{\mu,0}(t)
&=
	\frac{
		2 g^{2}
	}{
		\pi
	}
	\int_{0}^{\infty}
	\dd k
    \frac{
        k^{2}
    }{
        \ep_{k}^{3}
    }
    \sin
    \left(
        2
        \ep_{k}
        t
    \right).
\label{eq:LoschmidtRate}
\end{align}
This recovers the $ T \to 0 $ limit of \cref{eq:gamma}, using \cref{eq:EnvironmentCorr} for the environment correlation.  
%
\begin{figure*}[t!]
    \includegraphics[width=225pt, height=170pt]{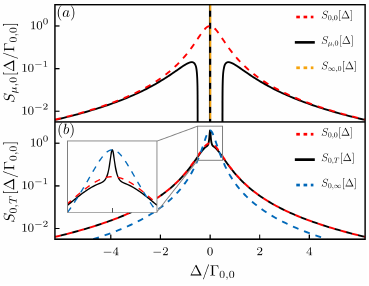}
    \includegraphics[width=130pt, height=180pt]{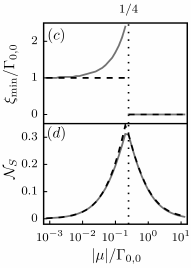}
    \includegraphics[width=130pt, height=180pt]{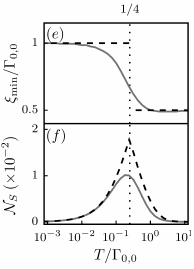}
    \vspace{-10pt}
    \caption{Steady-state emission spectrum of the probe qubit along the (a) gap and (b) temperature axes. The corresponding spectral measure $\mathcal{N}_{S}$ of the probe qubit calculated along the gap (d) and temperature (f) using the Lorentzian spectra at the three fixed points (dashed-black), with $\xi\in\qty{\rate{0, 0}, \rate{\infty, 0}, \rate{0, \infty} }$ in \cref{eq:specNMmeasure}, and an arbitrary Lorentzian profile (solid-grey) as reference Markovian spectra, respectively. Panels (c) and (e) show the zero-frequency widths, $\xi_{\rm min}$, that minimize the JSD used for calculating the $\mathcal{N}_{S}$ shown in panels (d) and (f), respectively.}
    \vspace{-10pt}
\label{fig:Spec}
\end{figure*}
%
%
\par
\textit{Spectral measure of RG flow.--} While single-time quantities such as the dephasing rate encode the environmental correlation, spectroscopic measurements of the probe permit a similar diagnosis in the steady-state or long-time limit. More pertinently, distinguishing probe non-Markovianity due to finite temperature versus finite gap at an arbitrary point in the TFIM phase diagram using only single-time quantities is challenging. In this section, we show how spectroscopic probe measurements can provide a powerful means to quantify and distinguish non-Markovianity induced by the RG flow along different scaling fields \cite{keefe2025}. To this end, we obtain the normalized emission spectrum of the qubit using the frequency-domain master equation construction introduced in ref.~\cite{keefe2025},
\begin{align}
	\spec{\mu,T}[\Delta]=
	\frac{1}{\pi}\Re \frac{1}{i\Delta-\corr{\mu, T}[\Delta]}
	\,,
\label{eq:SSSpectrum}
\end{align}
where $\Delta=\omega-\omega_{\tS}$ and $C_{\mu,T}[\Delta]$ denotes the environment correlation function in frequency domain. Working in the frequency domain  has the additional advantage of maintaining full time-nonlocality in the probe state and bypassing the need to explicitly calculate its two-time correlator (see \cref{app:SSEmissionSpectrum}). The real and imaginary parts of $C_{\mu,T}[\Delta]$ are given by 
\vspace{-10pt}
\begin{align}
    \Re \corr{\mu, T}[\Delta]
    &=
    -g^2\hspace{-3pt}\int_{|\mu|}^{\infty}
    \hspace{-3pt} \dd{\ep_k} 
    \sqrt{1-\frac{\mu^2}{\ep_k^2}}
    f_{k}(T)
    \delta(|\Delta|-2\epsilon_{k}),
    \nonumber\\
    \Im \corr{\mu, T}[\Delta]
    &=
    -g^2\mathcal{P}\int_{|\mu|}^{\infty} 
    \hspace{-3pt}\frac{\dd{\ep_k}}{\pi} 
    \sqrt{1-\frac{\mu^2}{\epsilon_{k}^2}}
   f_{k}(T)
    \frac{2\Delta}{\Delta^2-4\epsilon_{k}^2},
    \nonumber
\end{align}
where $f_{k}(T) = 2(1-2 \langle\Delta n_{k}(T)\rangle^{2})$, with $\langle\Delta n_{k}(T)\rangle^{2}$ denoting the number fluctuations associated with the $k^{\rm th}$ Majorana mode, defines a $T$-dependent weight factor which reduces with increase in $T$. As shown in \cref{fig:Spec}, the spectrum in \cref{eq:SSSpectrum} becomes exactly Lorentzian at each of the RG fixed points, with widths given by the corresponding rates shown in \cref{fig:Rates}. Interestingly, while both finite gap and finite temperature impose a finite IR cutoff on the TFIM dynamics, they lead to very distinct spectral signatures in the probe qubit. While a finite gap burns a hole in the qubit emission spectrum, leading to \emph{finite-frequency} spectral wings, a finite temperature leads to an additional \emph{zero-frequency} spectral feature that progressively broadens with increase in temperature. 

Besides providing a visual diagnostic for distinguishing relevant deformations, the probe spectrum also enables a quantitative comparison of non-Markovianity using the recently proposed spectral measure which quantifies the information cost of making a Markovian approximation based on comparing non-Markovian and Markovian spectra \cite{keefe2025}. 
Since there are three reference Markovian spectra for the probe qubit, corresponding to the three RG fixed points of the TFIM, we use a modified spectral measure defined as (see \cref{app:Quantifying non-Markovianity})
\begin{align}
    d_S(\mu, T; \xi)
    \equiv
    \sqrt{\jsd(S_{\mu, T}||  S_M^\xi)},
    \hspace{4pt}
    \mathcal{N}_{S} \equiv \min_\xi  d_S^{2} (\mu, T; \xi),
    \nonumber
\label{eq:specNMmeasure}
\end{align}
where $d_{S}$ is now the spectral distance of the probe spectrum to the nearest Markovian spectrum in the phase diagram, parametrized as $S_M^\xi[\Delta] = 
    (1/\pi) \xi/(\Delta^2+\xi^2)$. 
As shown in \cref{fig:Spec}, $\mathcal{N}_{S}$ sharply falls to zero as RG fixed points are approached. Moreover, away from the fixed points, a finite gap leads to a significantly stronger non-Markovianity as opposed to finite temperature, as reflected in the respective magnitudes of $\mathcal{N}_{S}$ calculated along the two axes. This aligns with the intuitive insight provided by the spectral measure of non-Markovianity that zero (finite)-frequency deformations of a Lorentzian correspond to weak (strong) non-Markovian corrections per unit bandwidth \cite{keefe2025}.
\par
Analogous to the `rate flow' in time, the `spectral flow' of the probe also emulates the TFIM RG flow between unstable and stable fixed points since 
$\spec{\mu, 0}[\Delta \to 0]=\spec{\infty, 0}[\Delta \to 0]=\Phi_0 \delta[\Delta]$, $\spec{\mu, 0}[\Delta \to \infty]=\spec{0, 0}[\Delta\to \infty]$, and $\spec{0, T}[0]=\spec{0, \infty}[0]$, $\spec{0, T}[\Delta\to\infty]=\spec{0, 0}[\Delta\to \infty]$
(see \cref{app:SSEmissionSpectrum}). We can further visualize the RG flow in the bulk of the phase diagram by performing a principal component analysis of features in the non-Markovian probe spectra obtained at any $(\mu, T)$ point in the TFIM phase diagram. To this end, we first map each $(\mu, T)$ point to a vector $ \boldsymbol{\mathcal{D}}(\mu,T) $ in ``spectral fixed-point space'' whose components represent the spectral distances to the three fixed points $\{\mu, T\} = \{0,0\}, \ \{0,\infty\}$, and $\{\infty,0\}$ (see \cref{app:FlowAnalysis}). In order to obtain a flat view of this curved surface, we \newline
(i) define a flow field by taking the logarithmic directional derivative of $\boldsymbol{\mathcal{D}}(\mu,T)$ along the diagonal direction in $ (\ln(\mu/\rate{0,0}), \ln(T/\rate{0,0})) $ space, 
\begin{align}
	\boldsymbol{\mathcal{F}}_{\mathbf{u}}(\mu, T)
=
	\left(
		\mathbf{u}
		\cdot
		\boldsymbol{\nabla}
	\right)
	\boldsymbol{\mathcal{D}}(\mu, T),
\end{align}
where $ \boldsymbol{\nabla} = (\partial_{\ln (\mu/\rate{0,0})}, \partial_{\ln (T/\rate{0,0})}) $ and $ \mathbf{u} $ is a vector in the $(\mu, T)$ plane, \newline
(ii) project it onto the plane spanned by the first two principal component directions \footnote{In this case, the third singular value is negligible compared to the first two.} obtained from the three fixed-point vectors, $ \mathcal{D}(0,0) $, $ \mathcal{D}(0,\infty) $, and $ \mathcal{D}(\infty,0)$.
\par
%
\begin{figure}[t!]
    \includegraphics[width=\columnwidth]{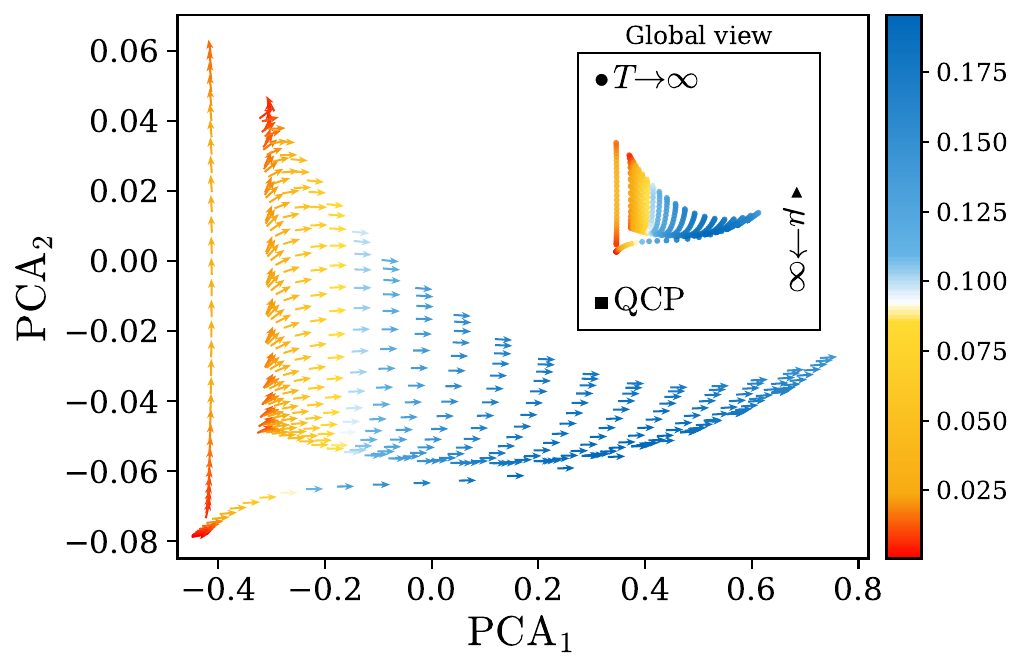}
    \vspace{-20pt}
    \caption{Directional derivative field $\boldsymbol{\mathcal{F}}_{\mathbf{u},2}(\mu, T)$ over a rectangular region in phase space spanned by ${\mu\in [0,0.4]}$, ${T\in [0,1.8]}$, while setting $g=1$. The inset shows the global structure with the fixed points. The flow in the selected region captures all the non-trivial trends with trivial extrapolation to the fixed points at infinities.} 
     \vspace{-10pt}
\label{fig:FlowPlot}
\end{figure}
%
\par
Since the beta functions $\beta_{\mu}\equiv d\mu/dl=\mu$, ${\beta_T\equiv dT/dl=T}$ are always linear for a non-interacting fermionic field theory such as the TFIM (see \cref{app:FlowAnalysis}), a natural choice is $\mathbf{u} = (1,1)/\sqrt{2} $. The resultant 2D flow field $\boldsymbol{\mathcal{F}}_{\mathbf{u},2} =(1/\sqrt{2})(d\boldsymbol{\mathcal{D}}_2/dl)$ maps the RG flow transformed by the Jacobian of the surface-to-surface map,
\begin{align}
    \frac{d\boldsymbol{\mathcal{D}}_2}{d\ell} = \begin{bmatrix} \partial_{\mu} \mathcal{D}_{2,1} & \partial_{T}\mathcal{D}_{2,1} \\ \partial_{\mu} \mathcal{D}_{2,2}& \partial_{T}\mathcal{D}_{2,2} \end{bmatrix} \begin{pmatrix} \beta_{\mu} \\ \beta_T \end{pmatrix} \equiv \mathbb{J}_{2} \bm{\beta} .
\label{eq:D2Jacob}
\end{align}
\Cref{fig:FlowPlot} shows the 2D  flow over a rectangular region of the phase diagram, with the flow along the $ \mu $-axis ($ T $-axis) shown along the lowest line of arrows (left-most line of arrows). The color of the arrows denotes the magnitude of the flow or change in non-Markovianity. How a linear flow $\bm{\beta}$  begets a highly inhomogeneous and anisotropic flow of $\boldsymbol{\mathcal{F}}_{\mathbf{u},2}$ through the bulk region can be inferred from the dominant eigenvalue of the Jacobian $\mathbb{J}_{2}$~\footnote{The overlapping vectors in $ \mathcal{F}_{\mathbf{u},2}(\mu, T) $ reflect the curvature of $ \mathcal{D}(\mu, T) $ in the full three-dimensional fixed-point space, where the construction of the Jacobian does not commute with the projection onto the principal plane.}: $\boldsymbol{\mathcal{F}}_{\mathbf{u},2}\approx |\lambda_{1} c_{1}| \mathbf{v}_1$, where $\mathbb{J}_{2}\mathbf{v}_1=\lambda_1\mathbf{v}_1$ with $c_1 = {\bf u \cdot v_{1}}$. Thus the direction and speed (color) of the flow field are set by $\mathbf{v_1}$ and $|c_{1}\lambda_1|$, respectively. We find that $ \boldsymbol{\mathcal{F}}_{\mathbf{u},2}(\mu, T) $ captures the off-axis flow away from the QCP, while emphasizing the overall flow toward the $ \mu \to \infty $ IR-stable fixed point in the bulk. This dominant flow to the $\mu \to \infty$ fixed point corroborates our earlier finding of rate flow towards $\Gamma_{\infty,0}$, even for an infinitesimal off-axis deformation in the TFIM phase diagram. 
%
\par
\textit{Conclusions.--}
Using the TFIM as an archetypal many-body environment coupled to a central qubit probe, we have established an explicit relationship between non-Markovian probe dynamics and the RG flow of TFIM. Specifically, we show how this can enable identification of fixed points via the scale-invariant Markovian rates they induce on the quantum probe, as well as a ``single-shot'' estimation of stability of fixed points and scaling dimensions (and/or critical exponents) of associated fields, via non-Markovian rate or spectral flow of the qubit induced by the TFIM environment at a single point in the $\{\mu, T\}$ phase diagram. The spectral flow can also distinguish the induced non-Markovianity due to different deformations about the fixed points, providing a means to faithfully map the RG flow in the TFIM phase diagram. 

There are several interesting questions that are natural extensions of our results: such as the effect of coupling modalities, both in terms of the effect of the scaling dimension of environment operator entering the system-environment interaction, as well as generalizations beyond the QND nature of the interaction and Gaussian fixed points considered here. In such cases, it will be interesting to deconvolve the effect of non-Gaussianity due to non-commuting interactions with the probe versus those due to non-Gaussian states of the many-body environment itself. It will also be interesting to explore extensions to quantum simulation platforms where advanced qubit readout capabilities can be leveraged to study new phases of engineered quantum matter, including those characterized by non-local order parameters \cite{Catuneanu2019} and anisotropic field theories \cite{Po2015}. 


\textbf{\textit{{Acknowledgments.--}}} The authors thank Ehud Altman, Aashish Clerk, Sebastian Diehl, and Andrew Higginbotham for useful comments and critiques; A.~K. acknowledges support from the Kavli Institute for Theoretical Physics (KITP) for hosting the long program on ``Many-body Quantum Optics" (NSF PHY-2309135) and the Thouless Institute of Quantum Matter at UW Seattle for hosting the workshop on ``Recent Advances in Open Systems" that facilitated many of these discussions. Discussions with Carlos Gonz\'alez-Guti\'errez on Loschmidt echo during the earlier stages of this work are gratefully acknowledged. This work was supported by the National Science Foundation under grants DMR-2047357 and DMR-2508447 and the Department of Energy under grants DE-SC0020360 and DE-SC0019461.


%


\appendix
\setcounter{secnumdepth}{2}

\section{Diagonalizing the TFIM Hamiltonian}
\label{app:diagTFIM}

In this appendix, we map the TFIM Hamiltonian in spin basis given by \cref{eq:TICHamiltonian} to a system of free fermions and diagonalize it \cite{Sachdev2011}. Using the standard Jordan-Wigner transformation,
\begin{subequations}
\begin{eqnarray}
    \hat{\sigma}_j^x & = & 1-2\hat{f}_j^\dag \hat{f}_{j} \, , \\
    \hat{\sigma}_j^z & = & -\prod_{i<j} \left( 1-2\hat{f}_j^\dag \hat{f}_{j} \right) \left( \hat{f}_i + \hat{f}_{i}^\dag \right) \, ,
\end{eqnarray}
\end{subequations}
we first write the Hamiltonian in \cref{eq:TICHamiltonian} as
\begin{align}
	\hH_{\tE}
=&
    -h\sum_{j=0}^N (1-2\hat{f}_j^\dag \hat{f}_{j})
    \nonumber \\
    &
   -J\sum_{j=0}^N \qty(
    \hat{f}_j^\dag \hat{f}_{j+1} 
    + 
    \hat{f}^\dag_{j+1}\hat{f}_j
    +
    \hat{f}_j^\dag \hat{f}^\dag_{j+1}
    +
    \hat{f}_{j+1} \hat{f}_{j}
    )
	\,.
\label{eq:FreeHamEnvFerm}
\end{align}

Next, we define the (discrete) Fourier transform
\begin{align}
	\hat{f}_n
&=
    \frac{1}{\sqrt{LN}}\sum_{m=-N/2}^{N/2-1}e^{ia n k_m }\hat{f}_{k_m} 
    \,,
    &
    \hspace*{10pt}
    k_m=\frac{2\pi m}{L}
	\,,
\label{eq:Fermions_Fourier}
\end{align}
to express this in momentum basis,
\begin{align}
	\hH_{\tE}
=
    \frac{1}{L}\sum_{m=-N/2}^{N/2-1} \Big[&2\qty(h-J\cos a k_m)\hat{f}^\dag_{k_m}\hat{f}_{k_m} 
    \nonumber \\
    &
    - iJ\sin a k_m\qty(\hat{f}^\dag_{k_m}\hat{f}^\dag_{-k_m}  + \hat{f}_{k_m}\hat{f}_{-k_m} ) \Big]
	\,.
\label{eq:FreHamEnvFermK}
\end{align}
Finally, we  diagonalize the above Hamiltonian by performing a Bogoliubov transformation,
\begin{align}
	\epsilon_k \gamma_k
=&
    \xi_k f_k - i r_k f_{-k}^\dag
	\,,
\label{eq:Bogoliubov}
\end{align}
to obtain the final form of the TFIM Hamiltonian,
\begin{align}
	\hH_{\tE}
=&
    \frac{1}{L}\sum_{m=-N/2}^{N/2-1} \ep_{k_m} \hgam_{k_m}\dagg\hgam_{k_m},
\label{eq:FreeHamEnvGammaDisc}
\end{align}
with the energy spectrum and Bogoliubov coefficients defined as
\begin{subequations}
\begin{align}
\label{eq:DispersionBare}
	\ep_{k}
&=
	2
	\sqrt{
		J^{2}
		+
		h^{2}
		-
		2
		h
		J
		\cos(a k)
	}
	\,, \\
    \xi_k
    &=
    2(h-J \cos ak)
    \,, \\
    r_k
    &=
    2J \sin ak
    \,.
\end{align}
\end{subequations}

We are interested in two field theory limits of this finite system. The first is the \textit{statistical field theory} (SFT) limit, in which we take the thermodynamic limit $ N \to \infty $ in final quantities holding the lattice spacing $ a $ fixed. This pushes $ L \to \infty $ and results in a field theory with a UV cutoff at $ \pi/a $. Lastly, we normal order the diagonalized Hamiltonian with respect to the environment state, which, for the quadratic Hamiltonian here, amounts to replacing $ \hH_{\tE} \to \mathop{:} \hH_{\tE} \mathop{:} \equiv \hH_{\tE} - \langle \hH_{\tE} \rangle_{\hrho_{E}} $. The resulting Hamiltonian is then given by
\begin{align}
	\hH_{\tE}
&=
    \int_{-\pi/a}^{\pi/a} \frac{\dd k}{2 \pi}
	\epsilon_{k}
    \mathop{:}
	\hgam_{k}\dagg
	\hgam_{k}
    \mathop{:}
	\,,
\label{eq:FreeHamEnvGamma}
\end{align}

The second limit we are interested in is the continuum limit $ a \to 0 $ of the SFT, that results in the \textit{quantum field theory} (QFT) of a massive fermion \cite{Sachdev2011}. In this limit, the Hamiltonian takes the same form as in \cref{eq:FreeHamEnvGamma}, however the expressions for the Bogoliubov coefficients simplify to
\begin{subequations}
\begin{align}
	\ep_{k} &= \sqrt{k^{2}+\mu^{2}} 
	\,,
\\
    \xi_k 
    &=
    \mu
    \,, \\
	r_{k} &= k
	\,,
\end{align}
\label{eq:ContinuumSubs}%
\end{subequations}
where we have defined the mass of the fermion field $ \mu = \frac{1}{a}  \big( 1 - \frac{h}{J} \big) $ and set the velocity $ v = 2Ja \equiv 1 $. Note that the quantum critical point of the TFIM maps to $ \mu = 0 $ in this limit. 

With the TFIM Hamiltonian diagonalized, we now express the interaction Hamiltonian in terms of the Majorana basis. Defining the environment subspace of the interaction as
\begin{align}
	\hat{\mathcal{O}}_{\tE}
    &\equiv
    \frac{g}{\sqrt{L}}
    \sum_{j=0}^{N}\sigma^x_j
	\,,
\label{eq:OESpin}
\end{align}
we have $\hH_{\text{int.}}=\hat{\sigma}^z_{\tS}\hat{\mathcal{O}}_E$. Applying the above transformations, we find
\begin{align}
	\mathop{:}
    \hat{\mathcal{O}}_{\tE}
    \mathop{:}
    =&
    -\frac{2g}{\sqrt{L}}\int_{-\pi/a}^{\pi/a}
    \frac{\dd{k}}{2\pi}
    \chi_k \mathop{:}\hgam_k\dagg \hgam_k \mathop{:}
    \nonumber \\
    &
    -\frac{2g}{\sqrt{L}}\int_{-\pi/a}^{\pi/a}
    \frac{\dd{k}}{2\pi}
    i r_k\qty(\hgam_k\dagg \hgam_{-k}\dagg + \hgam_k \hgam_{-k})
	\,,
\label{eq:OEMajoranaFull}
\end{align}
in the SFT limit. Here we have normal ordered the interaction to ensure that the resulting QME depends only on the connected correlation function. The first term, however, is still problematic, as it leads to an interaction term which is diagonal in the free Hamiltonian basis. This leads to a secular divergence in the effective system dynamics, thus invalidating the quasi-static bath and Markov approximations. We may instead include this first term in the free Hamiltonian, 
\begin{align}
    \hat{\mathcal{O}}_{\tE}
    \to&
    -\frac{2ig}{\sqrt{L}}\int_{-\pi/a}^{\pi/a}
    \frac{\dd{k}}{2\pi}
     r_k\qty(\hgam_k\dagg \hgam_{-k}\dagg + \hgam_k \hgam_{-k})
	\,, \nonumber \\
    \hH_{\text{free}}
    \to&
    \hH_{\text{free}}-\frac{2g}{\sqrt{L}}\int_{-\pi/a}^{\pi/a}
    \frac{\dd{k}}{2\pi}
    \chi_k \mathop{:}\hgam_k\dagg \hgam_k \mathop{:}
    \,.
\label{eq:OEMajoranaShifted}
\end{align}
Doing so results in a system-dependent shift on $\epsilon_k$, however, this will ultimately vanish in the thermodynamic limit $L\to\infty$.
%
\section{The environment corelation function and dephasing rate}
\label{app:CorrFunction}
The simplest qubit observable that can be used to probe the TFIM is the qubit dephasing rate. In this appendix, we present a derivation of the Redfield master equation obtained for the probe qubit upon tracing out the TFIM environment, including a calculation of the environment correlation function that appears in the master equation, given by \cref{eq:EnvironmentCorr}. 

We start by assuming that the qubit and TFIM are initially uncorrelated and that the interaction is turned on at the time $t_0$. The evolution of the composite density operator $ \jdoi(t) $ in the interaction picture is described by the von Neumann equation,
\begin{align}
	\frac{ \dd }{ \dd t }
	\jdoi(t)
=
	-
	i
	\left[
		\hH_{\ihip}(t)
		,
		\jdoi(t)
	\right]
	\,,
\label{eq:vNE}
\end{align}
where $ \hH_{\ihip}(t) = e^{i \hH_{0} (t-t_{0})} \hH_{\text{int}} e^{-i \hH_{0} (t-t_{0})} $, with $ \hH_{0} = \hH_{\tS} + \hH_{\tE} $, is the interaction Hamiltonian in the interaction picture. Following the standard derivation of the master equation \cite{Bowen:2024emo,keefe2025}, we integrate \cref{eq:vNE} with respect to time, substitute the resulting expression for $ \jdoi(t) $ on the right-hand side of \cref{eq:vNE}, and trace over the environment. 

Further, we work in the weak-coupling regime, taking the environment to be in a thermal steady-state at temperature $T$ ($ k_{B} = 1 $), given by
\begin{align}
\label{eq:ThermEnvDensityOp}
	\edoi
\to
	\edoi(T)
&=
	\frac{
		1
	}{
		Z
	}
	e^{
		-
		\hH_{\tE}/T
	}
	\,,
\end{align}
where $ Z = \trE e^{ - \hH_{\tE}/T } $, so $ \trE \edoi(T) = 1 $, and make the Born approximation $ \jdoi(t_{1}) \to \sdoi(t_{1}) \otimes \edo(T) $ under the integral. This yields the following equation for $\sdoi(t)$,
\begin{align}
	\frac{ 
		\dd 
		\sdoi
	}{ 
		\dd t 
	}
&=
	-
	\int_{t_{0}}^{t}
	\dd t_{1}
	\trE
	\left[
		\hH_{\ihip}(t)
		,
		\left[
			\hH_{\ihip}(t_{1})
			,
			\sdoi(t_{1})
			\edo(T)
		\right]
	\right]
	,
\label{eq:BornME}
\end{align}
where we have used that $ \hH_{\ihip} $ is normal ordered with respect to the environment, so that ${\trE \{ \hH_{\ihip}(t) \edo(T) \} = 0}$. Lastly, we assume that the environment correlation time is much shorter than that for the system, so that $ \sdoi(t_{1}) \approx \sdoi(t) $ over the domain of integration. Then, transforming to the Schr\"odinger picture gives Eq.~(\ref{eq:Redfieldequation}) of the main text, where we have switched from $ \sdoi $ to $ \sdo $ to distinguish the representations of density operator in interaction and Schr\"odinger pictures respectively.

We note that we have presented the thermodynamic limit of the environment operator $ \hat{\mathcal{O}}_{\tE} $ in \cref{eq:EnvironmentOperator}; however, to properly handle volume factors, here we will compute the correlation function in the large $ N $ limit, before taking $ N \to \infty $. Thus, we begin with the environment operator in the Schr\"odinger picture 
\begin{align}
	\hat{\mathcal{O}}_{\tE}
&=
    \frac{i g}{L^{3/2}}
    \sum_{m = -N/2}^{N/2-1}
	\frac{r_{k_{m}}}{\ep_{k_{m}}}
	\left(
    \hgam_{-k_{m}}\dagg\hgam_{k_{m}}\dagg-\hgam_{k_{m}}\hgam_{-k_{m}}
	\right) \,,
\label{eq:EnvironmentOperatorApp}
\end{align}
In the interaction picture, the time dependence of $ \hat{\mathcal{O}}_{\tE}(t) $ is entirely from $ \hgam_{k_{m}}(t) $ and $ \hgam_{k_{m}}\dagg(t) $ given by 
\begin{align}
    \hgam_{k_{m}}(t)
&=
    e^{- i\ep_{k_{m}}\left( t -  t_{0}\right)}\hgam_{k_{m}}
    \,.
\end{align}
Therefore, the environmental correlator can be written as
\begin{align}
\label{eq:CorrDisc}
&
	\corr{\mu,T}(t,t_{1})
=
	\frac{g^{2}}{L^{3}}
    \sum_{m,n = -N/2}^{N/2-1}
	\frac{r_{k_{m}}r_{k_{n}}}{\ep_{k_{m}}\ep_{k_{n}}}
\nn
&\hspace{10pt}
    \times
	\bigg[
		\langle
			\hgam_{k_{m}}
			\hgam_{-k_{m}}
			\hgam_{-k_{n}}\dagg
			\hgam_{k_{n}}\dagg
		\rangle_{T}
		e^{i2 [\ep_{k_{n}}\left(t_{1}-t_{0}\right)
        -\ep_{k_{m}}\left(t-t_{0}\right)]}
	\nn
	&\hspace{25pt}
		+
		\langle\hgam_{-k_{m}}\dagg\hgam_{k_{m}}\dagg\hgam_{k_{n}}\hgam_{-k_{n}}\rangle_{T}
		e^{i2[\ep_{k_{m}}\left(t-t_{0}\right)
        -\ep_{k_{n}}\left(t_{1}-t_{0}\right)]}
	\bigg]
	\,,
\end{align}
where we used the shorthand $ \langle \bullet \rangle_{T} = \trE \{ \bullet \edoi(T) \} $ and that the thermal state is diagonal to evaluate the vanishing terms with an imbalance between creation and annihilation operators. The two remaining four-point expectation values can be computed with Wick's contractions, implemented for fermions through repeated application of the relation
\begin{align}
	\langle
		\hat{f}_{0}
		\hat{f}_{1}
		\cdots
		\hat{f}_{2n}
	\rangle
&=
	\sum_{i=1}^{2n}
	(-1)^{i+1}
	\langle
		\hat{f}_{0}
		\hat{f}_{i}
	\rangle
	\left\langle
		\cdots
		\hat{f}_{i-1}
		\hat{f}_{i+1}
		\cdots
	\right\rangle
	\,.
\end{align}
Then it is straightforward to calculate the necessary two-points, finding
\begin{subequations}
\begin{align}
\label{eq:ThermalgamgamdApp}
	\langle
		\hgam_{k_{m}}
		\hgam_{k_{n}}\dagg
	\rangle_{T}
&=
	L
	\delta_{m,n}
	\frac{e^{\frac{\ep_{k_{n}}}{2 T}}}
    {2\cosh\left[\frac{\ep_{k_{n}}}{2 T}\right]}
	\,,
\\
\label{eq:ThermalgamdgamApp}
	\langle
		\hgam_{k_{m}}\dagg
		\hgam_{k_{n}}
	\rangle_{T}
&=
	L
	\delta_{m,n}
	\frac{
		e^{
			-
			\frac{
				\ep_{k_{n}}
			}{
				2 T
			}
		}
	}{
		2
		\cosh
		\left[
			\frac{
				\ep_{k_{n}}
			}{
				2 T
			}
		\right]
	}
	\,,
\end{align}
\end{subequations}
leaving us with \footnote{Had we directly used \cref{eq:EnvironmentOperator} to compute the correlation function, we would simply have to use that, in the thermodynamic limit, $ L \delta_{mn}|_{n=m} \to 2 \pi \delta(k_{m}-k_{n})|_{n=m} = 2 \pi \delta(0) $, to obtain the same result.}
\begin{align}
\label{eq:FiniteEnvironmentCorr}
	\corr{\mu,T}(t,t_{1})
&=
    \frac{4 g^{2}}{\pi}
    \int_{0}^{\frac{\pi}{a}}
    \dd k
	\frac{
		r_{k}^{2}
	}{
		\ep_{k}^{2}
	}
	\frac{
		\cos
		[
			2
			\ep_{k}
			(
				t
				-
				t_{1}
				+
				\frac{
                    i
				}{
					2
                    T
				}
			)
		]
	}{
		1
		+
		\cosh
		\left[
			\ep_{k}/T
		\right]
	}
	\,,
\end{align}
where we evaluated the summation over $ m $, then took the thermodynamic limit, dropping the indices since $ k_{n} $ becomes a continuous variable, and used that the integrand is even in $ k $ to write the limits $ 0 $ to $ \pi/a $.

Using \cref{eq:ContinuumSubs}, we take the QFT limit of \cref{eq:FiniteEnvironmentCorr}. Then making the change of variables $ k = (\ep_{k}^{2} - \mu^{2})^{1/2} \implies \dd k = (\ep_{k}^{2} - \mu^{2})^{-1/2} \ep_{k} \dd \ep_{k} $, we find
\begin{align}
	\corr{\mu,T}(t,t_{1})
&=
	\frac{4 g^{2}}{\pi}
    \hspace{-2pt}
	\int_{|\mu|}^{\infty}
    \hspace{-2pt}
	\dd \ep_{k}
	\sqrt{
		1
		-
		\frac{
			\mu^{2}
		}{
			\ep_{k}^{2} 
		}
	}
	\frac{
		\cos
		[
			2
			\ep_{k}
			(
				t
				-
				t_{1}
				+
				\frac{
    				i
				}{
					2 T
				}
			)
		]
	}{
		1
		+
		\cosh
		\left[
			\ep_{k}/T
		\right]
	}
	.
\label{eq:EnvironmentCorrApp}
\end{align}
This matches \cref{eq:EnvironmentCorr}. 

\subsection{Zero temperature}
\label{app:RateZeroTemp}
At zero temperature with a finite gap, \cref{eq:EnvironmentCorrApp} becomes
\begin{align}
\label{eq:FiniteEnvironmentCorrvac}
	\corr{\mu,0}(t,t_{1})
&=
	\frac{4 g^{2}}{\pi}
    \hspace{-2pt}
	\int_{|\mu|}^{\infty}
    \hspace{-2pt}
	\dd \ep_{k}
	\sqrt{
		1
		-
		\frac{
			\mu^{2}
		}{
			\ep_{k}^{2} 
		}
	}
    e^{
        -
        i
        2
        \ep_{k}
        (
            t
            -
            t_{1}
        )
    }
	\,.
\end{align}
At the quantum critical point, $ \mu = 0 $, \cref{eq:FiniteEnvironmentCorrvac} can be evaluated using the Sokhotski-Plemelj theorem, from which one can obtain the identity
\begin{align}
	\int_{0}^{\infty}
	\dd k
	e^{
		-
		i
		k
        t
	}
=
    \pi
	\delta
	\left(
		t
	\right)
	-
    i
	P
	\Big(
		\frac{
			1
		}{
			t
		}
	\Big)
    \,,
\label{eq:SokPlemId}
\end{align}
where $ P $ indicates the Cauchy principal value. We find
\begin{align}
\label{eq:CriticalCorrVacuum}
	\corr{0,0}(t,t_{1})
=
    2
    g^{2}
	\delta
	\left(
		t
		-
		t_{1}
	\right)
	-
	\frac{
		i
        2
        g^{2}
	}{
		\pi
	}
	P
	\Big(
		\frac{
			1
		}{
			t
			-
			t_{1}
		}
	\Big)
	\,,
\end{align}
Using this in \cref{eq:gamma}, and multiplying by a factor of $ 1/2 $ since the delta function evaluates the $ t_{1} $ integral at the upper limit $ t $, we obtain the vacuum critical point decoherence rate
\begin{align}
	\rate{0,0}
&=
	g^{2}
	\,.
\end{align}

We now consider the dephasing rate for $ \mu \neq 0 $. Using the real part of \cref{eq:EnvironmentCorrApp} in \cref{eq:gamma} and computing the $ t_{1} $ integral, we have
\begin{align}
	\rate{\mu,0}(t,t_{1})
&=
	\frac{2 g^{2}}{\pi}
	\int_{0}^{\infty}
    \hspace{-2pt}
	\dd \ep_{k}
    \theta( \ep_{k} - |\mu| )
	\sqrt{
		1
		-
		\frac{
			\mu^{2}
		}{
			\ep_{k}^{2} 
		}
	}
    \frac{
        \sin
        \left[
            2
            \ep_{k}
            t
        \right]
    }{
        \ep_{k}
    }
	\,,
\label{eq:FiniteEnvironmentRatevac}
\end{align}
where we have set $ t_{0} = 0 $ for simplicity (this is recovered by letting $ t \to t-t_{0} $). We can now compute \cref{eq:FiniteEnvironmentRatevac} as a Fourier sine transform using \texttt{Mathematica}, we have
\begin{align}
\label{eq:finiteMuRatevac}
	\rate{\mu,0}(t)
&=
	g^{2} 
	J_{0}
	\left(
		2 
		|\mu|
		t
	\right)
	-
	2 
	g^{2} 
	|\mu|
	t  
	\big(
		1
		+
		J_{1}
		\left(
			2 
			|\mu|
			t
		\right)
	\big)
\nn
&\hspace{12pt}
	+
	2
	\pi
	g^{2} 
	|\mu|^{2}
	t^2
	\Big(
		\Big[
			\frac{2}{\pi}
			-
			\pmb{H}_{1}
                \hspace{-2pt}
			\left(
				2 
				|\mu|
				t
			\right)
		\Big] 
		J_{0}
		\left(
			2 
			|\mu|
			t 
		\right)
\nn
&\hspace{80pt}
		+
		\pmb{H}_{0}
            \hspace{-2pt}
		\left(
			2 
			|\mu|
			t 
		\right)
		J_{1}
		\left(
			2 
			|\mu|
			t 
		\right)
	\Big)
	\,,
\end{align}
where $ J_{n} $ is the Bessel function of the first kind and $ \pmb{H}_{n} $ is the Struve function. In the limit $ \mu \to 0 $, the first term recovers the quantum critical rate and the others vanish.

The behavior at finite $\mu$ can be seen as non-Markovian distortions over this critical Markovian behavior, as becomes explicit by looking at
\begin{equation}
\begin{aligned}
\Re[C_{\mu,0}(t)]&= -2g^2|\mu| \left[ 1 + J_1(2|\mu|t)\left(1 - \pi|\mu|t \mathbf{H}_0(2|\mu|t)\right)\right. \\
&\left.+ |\mu|t J_0(2|\mu|t)\left(-2 + \pi \mathbf{H}_1(2|\mu|t)\right) \right]+2g^2\delta(t).  
\end{aligned}
\label{eq:finiteMuCorrVac}
\end{equation}
Appearance of the Markovian piece $\delta(t)$ can be traced back to \cref{eq:EnvironmentCorr}, when $\mu\to 0$. The remaining part $\propto |\mu|$, is the result of subtracting the corresponding piece from the integrand.
That $\Re[C_{\mu,0}(t)]$ can't be found entirely as a continuous function, and a distribution function piece as $\sim \delta(t)$ is mandated directly follows from \cref{eq:gamma}, since the latter can't beget a finite dephasing rate as $t\to 0$, if $\Re[C_{\mu,0}(t)]$ is continuous at $t=0$. 
%
\subsection{Finite temperature}
\label{app:RateFiniteTemp}
%
For finite temperature, we can obtain analytical expressions at the critical point $ \mu = 0 $. In this case, \cref{eq:EnvironmentCorrApp} becomes
\begin{align}
\label{eq:CPEnvironmentCorrT}
	\corr{0,T}(t,t_{1})
&=
	\frac{4 g^{2}}{\pi}
    \hspace{-2pt}
	\int_{0}^{\infty}
    \hspace{-2pt}
	\dd k
	\frac{
		\cos
		[
			2
			k
			(
				t
				-
				t_{1}
				+
				\frac{
    				i
				}{
					2 T
				}
			)
		]
	}{
		1
		+
		\cosh
		\left[
			k/T
		\right]
	}
    \,.
\end{align}
Since the rate only depends on the real part of the correlation function, we will not consider the imaginary part. We then further separate the real part of the correlation function into zero and finite temperature contributions as  
\begin{align}
\label{eq:RealPartCorr}
    \Re[\corr{0,T}(t,t_{1})] 
= 
    \Re[\corr{0,0}(t,t_{1})] 
    + 
    \mathcal{C}_{T}(t,t_{1})
    \,, 
\end{align}
where
\begin{align}
\label{eq:CPEnvironmentCorrInfiniteT}
    \Re[\corr{0,0}(t,t_{1})]
&=
	\frac{4 g^{2}}{\pi}
    \hspace{-2pt}
	\int_{0}^{\infty}
    \hspace{-2pt}
	\dd k
    \cos
    [
        2
        k
        (
            t
            -
            t_{1}
        )
    ]
\nn
&=
    2
	g^{2}
    \delta
    (
        t
        -
        t_{1}
    )
    \,,
\\
\label{eq:CPEnvironmentCorrfiniteTint}
    \mathcal{C}_{T}(t,t_{1})
&=
    -
	\frac{2 g^{2}}{\pi}
    \hspace{-2pt}
	\int_{0}^{\infty}
    \hspace{-2pt}
	\dd k \, \langle|\Delta n_{k}|\rangle^{2}\cos[2k(t -t_{1})]
    \,.
\end{align}
with
\begin{equation}
    \langle\Delta n_{k} \rangle = \left(\langle n_{k}^{2}\rangle - \langle n_{k}\rangle^{2}\right)^{1/2}=\frac{1}{2}\sech\left(\frac{k}{2 T}\right),
\end{equation}
denoting the number fluctuations associated with the $k^{\rm th}$ Majorana mode.
Note that for the quantum critical point at $ T = 0 $, $ \mathcal{C}_{0}(t,t_{1}) = 0 $ and \cref{eq:RealPartCorr} reduces to the real part of \cref{eq:CriticalCorrVacuum}; for the infinite temperature limit, $ \mathcal{C}_{\infty}(t,t_{1}) = - g^{2} \delta(t - t_{1}) $, and as $ T \to \infty $, the real part of the environment correlation function becomes
\begin{align}
\label{eq:CriticalCorrInfTemp}
	\corr{0,\infty}(t,t_{1})
&=
	g^{2}
	\delta(t-t_{1})
	\,.
\end{align}
For a general temperature, \cref{eq:CPEnvironmentCorrfiniteTint} can be computed using the Fourier cosine transform in \texttt{Mathematica}, we find
\begin{align}
\label{eq:CPEnvironmentCorrfiniteT}
    \mathcal{C}_{T}(t,t_{1})
&=
    -   
	8 
    g^{2} 
    T^{2} 
    (t-t_{1})
    \hspace{1pt}
    \text{csch}
    \hspace{-2pt}
    \left[
    	2 \pi T (t-t_{1})
    \right]
    .
\end{align}
Using that 
\begin{align}
    \lim_{T \to \infty}
	8 
    T^{2} 
    (t-t_{1})
    \hspace{1pt}
    \text{csch}
    \hspace{-2pt}
    \left[
    	2 \pi T (t-t_{1})
    \right]
&=
    \delta(t-t_{1})
    \,,
\end{align}
which can be verified through its action on a test function $ \phi(t) $:
\begin{eqnarray}
    \lim_{T \to \infty}
    \int_{-\infty}^{\infty}
    \dd t
    T^{2}
    t
    \hspace{1pt}
    \text{csch}
    \left[
    	2 \pi T t
    \right]
    \phi(t)
& = &
    \frac{\phi(0)}{4 \pi^{2}}
    \int_{-\infty}^{\infty}
    \dd u
    \frac{u}{\sinh(u)} \nonumber \\
& = &  
    \frac{ \phi(0) }{8}
    ,
\end{eqnarray}
where we have assumed that the limit and integral can be exchanged, we again find that $ \mathcal{C}_{\infty}(t,t_{1}) = - g^{2} \delta(t - t_{1})  $. Thus, \cref{eq:RealPartCorr} is
\begin{align}
    \Re[\corr{0,T}(t,t_{1})] 
&=
    2
    g^{2}
    \delta(t-t_{1})
\nn
&\hspace{11.5pt}
    -   
	8 
    g^{2} 
    T^{2} 
    (t-t_{1})
    \hspace{1pt}
    \text{csch}
    \hspace{-2pt}
    \left[
    	2 \pi T (t-t_{1})
    \right]
    \,,
\end{align}
using this in \cref{eq:gamma} and computing the $ t_{1} $ integral, we find the temperature dependent dephasing rate to be
\begin{align}
\label{eq:CPRateT}
	\rate{0,T}(t)
&=
    \frac{
        g^{2}
    }{
        2
    }
    +
    \frac{
        4
        g^{2}
        T
        (t-t_{0})
    }{
        \pi
    }
    \ln
    \big(
        \coth
        [
            \pi
            T
            (t-t_{0})
        ]
    \big)
\nn
&\hspace{11.5pt}
    +
    \frac{
        g^{2}
    }{
        \pi^{2}
    }
    \Big(
        4
        \text{Li}_{2}
        \big[
            e^{
                -
                2
                \pi
                T
                (t-t_{0})
            }
        \big]
        -
        \text{Li}_{2}
        \big[
            e^{
                -
                4
                \pi
                T
                (t-t_{0})
            }
        \big]
    \Big)
	\,,
\end{align}
where 
\begin{align}
	\text{Li}_{2}(x) 
\equiv
	\sum_{k = 1}^{\infty}
	\frac{x^{k}}{k^{2}}
	\,,
\end{align}
is the dilogarithm function.
%
\subsection{Short time dephasing rate at finite gap and temperature}
\label{app:RateFiniteGapTemp}
%
Let us work after setting $t_0=0$ and $t\to t-t_0$. From \cref{eq:EnvironmentCorrApp},
\begin{align}
& \text{Re}[C_{\mu,T}(t)] = \text{Re}[C_{\mu,0}(t)]+ C_{\text{mix}}(t), \label{eq:CorrMuT1}\\
& C_{\text{mix}}(t)=-\frac{8g^2}{\pi} \int_{|\mu|}^\infty d\epsilon_k 
	\sqrt{1-\frac{\mu^{2}}{\ep_{k}^{2}}}  \langle \Delta n_{k} \rangle^{2}\cos(2\epsilon_k t), \nonumber \\ \label{eq:CorrMuT2}
\end{align}
where the first term is given by \cref{eq:finiteMuCorrVac}, and the $C_{\text{mix}}$ becomes same as $\mathcal{C}_{T}$ when evaluated at $\mu=0$.
\par
At such a generic location in the parameter space, both $\tau_{\mu}\equiv(2|\mu|)^{-1}$ and $\tau_{T}\equiv\pi (8T)^{-1}$ appear as competing timescales in the correlation function, with their interplay giving rise to a smooth crossover transition as the TFIM traverses between the $\mu=0$ and $T=0$ axes. 
\par
In the presence of two competing timescales, the short time regime is determined by ${\rm min}\{\tau_{\mu}, \tau_{T}\}$. This can be directly seen from the two pieces in \cref{eq:CorrMuT1}. The zero-temperature contribution of the first term in the dephasing rate has time-dependence only appearing in combination of $t/\tau_{\mu}$, as found in \cref{eq:finiteMuRatevac}. Hence, it can be approximated by expanding to low orders of $t$ when $t\ll \tau_{\mu}$. 
\par
For $C_{\text{mix}}(t)$ defined in \cref{eq:CorrMuT2}, we can expand the $\cos(2\epsilon_kt)$ factor inside the integral to low orders in $t$ with the order set by $\mathcal{O}(t)\ll \mathcal{O}(1/T)$; this is because $\mathcal{C}_{\text{mix}}$ is an analytic function of $t$ and the thermal occupation factor $n_{k}^{2}$ exponentially suppresses contributions for $\epsilon_k>\mathcal{O}(1/T)$ in the integral. Using this in \cref{eq:gamma}, we obtain the linear-order expansion of the dephasing rate in the TFIM bulk,
\begin{align}
&\Gamma_{\mu,T}(t)\approx g^2\left(1-2|\mu|t\right)-\mathcal{A}_{\text{mix}}t+\mathcal{O}(t^2),\label{eq:ShortTimeRate}\\
&\mathcal{A}_{\text{mix}}= \frac{2g^2|\mu|}{\pi } \int_{1}^\infty dx 
\sqrt{1 - \frac{1}{x^{2}}}\sech^2\left(\frac{|\mu| x}{2T}\right),
\end{align}
where the first term is consistent with \cref{eq:ShortTimeRatesCrit}, and the second term follows from the $\mathcal{O}(t^0)$ expansion of $\mathcal{C}_{\text{mix}}$. Even though $\mathcal{A}_{\text{mix}}$ does not admit a close form expression, we can use the asymptotic expansion of the secant term which, terminated after $\mathcal{O}(10)$ terms, gives a reasonable approximation for small to moderate temperatures ${T\leq\mathcal{O}(|\mu|)}$,
\begin{align}
\mathcal{A}_{\text{mix}} &= \frac{8g^2|\mu|}{\pi} \sum_{n=1}^\infty (-1)^{n-1} n \int_{1}^\infty dx \sqrt{1 - \frac{1}{x^{2}}} e^{-z_n x} \nonumber\\
&= \frac{8g^2|\mu|}{\pi} \sum_{n=1}^\infty (-1)^{n-1} n \int_{z_n}^\infty \frac{K_1(q)}{q} dq,\\
\int_{z_n}^\infty \frac{K_1(q)}{q} dq&=K_1(z_n) - \frac{\pi}{2} + z_n K_0(z_n) \nonumber\\
&~~~~+\frac{\pi z_n}{2} \left[ K_0(z_n) \mathbf{L}_1(z_n) + K_1(z_n) \mathbf{L}_0(z_n) \right],
\label{eq:Amix}
\end{align}
where $z_n=n|\mu|/T$.
\par
Using eq. (\ref{eq:ShortTimeRate}), we can define the correlation time valid at a generic ($\mu,T$) point in TFIM phase diagram as,
\begin{equation}
\tau_B\equiv \frac{1}{2|\mu|+\mathcal{A}_{\text{mix}}/g^2},
\end{equation}
It can be observed from \cref{eq:ShortTimeRate}, that the ratio $\tau_B/\tau_T$ only depends on the ratio $z_t\equiv \tau_T/\tau_{\mu}$, and as such assumes the form of a scaling function. Hence, we track the interpolation of the time-constant across the crossover regime, through an emergent exponent $\eta_t$ describing the mixing of $\tau_T$ and $\tau_{\mu}$, defined as,
\begin{align}
&\frac{\tau_B}{\tau_T}=z_t^{\eta_t}\implies \tau_B=\tau_{\mu}^{-\eta_t}\tau_T^{1+\eta_t}.
\end{align}
The exponent $\eta_t$ is only a function of the ratio of the energy scales $z\equiv|\mu|/T=4z_t/\pi$, and approaches zero as $z\to 0$, signifying the scaling dimension $D_T=1$ along the zero-gap axis. For the opposite limit, $z\to\infty$, $\eta_t\to -1$ corresponding to $D_{\mu}=1$. This crossover through the finite $z$ regime is shown in \cref{fig:Rates}(c).

Inferring scaling dimension from bath time-constant follows from exploiting the equivalence of coarse graining between spatial and temporal scales in a Lorentz-invariant theory, as $\tau_{B} \rightarrow \tau_{B} b^{-1}$, where $b$ is an effective `blocking time'.  The relationship between the scaling fields $\{\mu, T\}$ and $\tau_{B}$, therefore, directly gives access to their respective scaling dimensions.

\section{Connection to CFT correlation functions}
\label{app:CFTCorrFunction}
A special property of the TFIM is that, at the critical point, it maps to a CFT with central charge $ c = 1/2 $, in which the real space correlation functions for this CFT are well-known power laws. In this appendix, we obtain the environment correlation function at criticality using the results from CFT, and show that in the zero and infinite temperature limits, the environment is delta correlated. 

Recall that, in the continuum limit, the TFIM maps to a free massive Majorana fermion given by the action~\cite{Itzykson:1989sx}
\begin{align}
	S_{\tE}
&=
	\int
	\dd z
	\dd \bar{z}
	\bigg[
		\bar{\psi}
		\partial 
		\bar{\psi}
		+ 
		\psi 
		\bar{\partial} 
		\psi 
		+
		i 
		\mu
		\bar{\psi}
		\psi
	\bigg]
	,
\label{eq:S_EMajorana}
\end{align}
where $ \psi $ and $ \bar{\psi} $ are real Grassman fields with ${\rm dim}[\psi] = {\rm dim}[\bar{\psi}] = 1/2$\, and
\begin{subequations}
\begin{align}
	z 
= 
	y 
	+ 
	i 
	x
	,
&\hspace{11.5pt}
	\partial 
= 
	\frac{1}{2}
	\left(
		\partial_{y}
		- 
		i
		\partial_{x}
	\right)
	,
\\
	\bar{z} 
= 
	y 
	- 
	i 
	x
	,
&\hspace{11.5pt}
	\bar{\partial}
= 
	\frac{1}{2}
	\left(
		\partial_{y}
		+ 
		i
		\partial_{x}
	\right)
	,
\end{align}
\end{subequations}
with $ y = - i t $. We can quantize this theory by imposing the anticommutation relations
\begin{subequations}
\begin{align}
	\left\{
		\hat{\psi}(x)
		,
		\hat{\psi}(x_{1})
	\right\}
&=
	\left\{
		\hat{\bar{\psi}}(x)
		,
		\hat{\bar{\psi}}(x_{1})
	\right\}
=
	\delta
	\left(
		x
		-
		x_{1}
	\right)
	\,,
\\
	\left\{
		\hat{\bar{\psi}}(x)
		,
		\hat{\psi}(x_{1})
	\right\}
&=
	0
	\,.
\end{align}
\end{subequations}
Under this mapping, the environment operator becomes
\begin{align}
    \hat{\mathcal{O}}_{\tE}(t)
=
	i
    \frac{2g}{\sqrt{L}}
    \int
    \dd x
    \hat{\bar{\psi}}(z,\bar{z})        			
    \hat{\psi}(z,\bar{z})
	\,.
\end{align}
Now, the environment correlation function can be written as 
\begin{align}
	\corr{\mu,T}(t,t_{1})
    & = 2
	\trE
	\big\{
		\hat{\mathcal{O}}_{\tE}\dagg(t)
		\hat{\mathcal{O}}_{\tE}(t_{1})
		\edo(T)
	\big\}
	\,\nonumber\\
&=
	\lim_{L \to \infty}
	\frac{2 g^{2}}{L}
    \int
	\dd x
	\dd x_{1}
	\Delta_{\mu,T}(z, \bar{z}, z_{1}, \bar{z}_{1})
	\,,
\label{eq:Crealsp}
\end{align}
where the spatial integrals are from $ - L/2 $ to $ L/2 $, and we have defined the real-space correlation function
\begin{align}
	\Delta_{\mu,T}(z, \bar{z}, z_{1}, \bar{z}_{1})
&=
	4
	\langle
		\psi(z,\bar{z})
		\bar{\psi}(z,\bar{z})
		\bar{\psi}(z_{1},\bar{z}_{1})
		\psi(z_{1},\bar{z}_{1})
	\rangle_{T}
	\,,
\end{align}
while suppressing the hats on the operators for notational simplicity. Now, using Wick's theorem in conjunction with the well-known two-point conformal correlators at the quantum critical point, ${\mu = T = 0}$ ~\cite{DiFrancesco:1997nk}
\begin{subequations}
\begin{align}
	\langle
		\psi(z,\bar{z})
		\psi(z_{1},\bar{z}_{1})
	\rangle_{0}
&=
	\frac{1}{2 \pi}
	\frac{
		1
	}{
		z
		-
		z_{1}
	}
    \,,
\\
	\langle
		\bar{\psi}(z,\bar{z})
		\bar{\psi}(z_{1},\bar{z}_{1})
	\rangle_{0}
&=
	\frac{1}{2 \pi}
	\frac{
		1
	}{
		\bar{z}
		-
		\bar{z}_{1}
	}
    \,,
\\
	\langle
		\psi(z,\bar{z})
		\bar{\psi}(z_{1},\bar{z}_{1})
	\rangle_{0}
&=
	0
    ,
\end{align}
\end{subequations}
leads to
\begin{align}
	\Delta_{0,0}(x, t, x_{1}, t_{1})
&=
	\frac{1}{\pi^{2}}
	\frac{
		1
	}{
		\left(
			x
			-
			x_{1}
		\right)^{2} 
		-
		\left(
			t
			-
			t_{1}
		\right)^{2} 
	}
	,
\label{eq:VacuumCFTCorr}
\end{align}
where $ z = -i t + i x $ and $ z_{1} = - i t_{1} + i x_{1} $. Using this in \cref{eq:Crealsp}, the environment correlation function can be computed
\begin{align}
	\corr{0,0}(t,t_{1})
&=
	\lim_{L \to \infty}
	\frac{2g^{2}}{\pi^{2} L}
	\int_{-\frac{L}{2}}^{\frac{L}{2}}
	\dd x
	\int_{-\frac{L}{2}}^{\frac{L}{2}}
	\dd x_{1}
\nn
&\hspace{11.5pt}
    \times
	\frac{
		1
	}{
		\left(
			x
			-
			x_{1}
		\right)^{2} 
		-
		\left(
			t
			-
			t_{1}
		\right)^{2} 
	}
	\,.
\end{align}
We regulate the limit $ t = t_{1} $ by pushing $ t - t_{1} \to t - t_{1} - i \epsilon $, compute the spatial integrals, and take the limit $ \ep \to 0 $, we find
\begin{align}
	\corr{0,0}(t,t_{1})
&=
	-
	\frac{i 2 g^{2}}{\pi}
	\lim_{\ep \to 0}
	\frac{
		1
	}{
		t
		-
		t_{1}
		-
		i
		\ep
	}
\nn
&=
    2 
    g^{2}
	\delta
	\left(
		t
		-
		t_{1}
	\right)
	-
	\frac{
		i
        2
        g^{2}
	}{
		\pi
	}
	P
	\Big(
		\frac{
			1
		}{
			t
			-
			t_{1}
		}
	\Big)
	\,,
\end{align}
which matches with the master equation correlation function, \cref{eq:CriticalCorrVacuum}.

The correlation function at finite temperature can be obtain from the vacuum case through the general mapping~\cite{Sachdev2011}
\begin{align}
	(it)
	\pm
	i
	x
\to
	\frac{1}{\pi T}
	\sin
	\left(
		\pi
		T
		\left(
			i
			t
			\pm
			i
			x
		\right)
	\right)
	\,.
\end{align}
Thus, from vacuum CFT case, \cref{eq:VacuumCFTCorr}, we obtain the finite temperature real-space correlation function
\begin{align}
	\Delta_{0,T}(x, t, x_{1}, t_{1})
&=
	T^{2}
	\text{csch}
	\Big(
		\pi T
		\left[
			(
				x
				-
				x'
			)
            -
    		(
                t
                -
                t_{1}
            )
		\right]
	\Big)
\nn
&\hspace{11.5pt}
    \times
	\text{csch}
	\Big(
		\pi T
		\left[
			(
				x
				-
				x'
			)
            +
    		(
                t
                -
                t_{1}
            )
		\right]
	\Big)
	.
\label{eq:FiniteTempCFTCorr}
\end{align}
We can then use this in \cref{eq:Crealsp} to obtain $ C_{0,T}(t,t_{1}) $. However, since $ \text{csch}(x) \to \infty $ as $ x \to 0 $, we need to regulate the spatial integrals in \cref{eq:Crealsp} carefully.

We consider the Taylor expansion for $ \text{csch}(x) $,
\begin{align}
	\text{csch}(x)
=	
	\sum_{n = -1}^{\infty}
	b_{n}
	x^{2n+1}
=	
	\frac{1}{x}
	+
	\sum_{n = 0}^{\infty}
	b_{n}
	x^{2n+1}
	\,,
\end{align}
where
\begin{align}
	b_{n}
=
	\frac{
		2
		\left(
			2^{2n+1}
			-
			1
		\right)
		B_{2n+2}
	}{
		(2n+2)!
	}
	\,,
\end{align}
with $ B_{n} $ being the $\text{n}^{\text{th}} $ Bernoulli number. The $ n = -1 $ term is $ 1/x $, and is the source of the divergence we need to regulate. Thus, we push the pole of the $ n = -1 $ term into the upper-half \footnote{This ensures that a positively oriented contour picks up the pole.} of the complex plane by taking $ x \to x - i \ep $, so 
\begin{align}
	\text{csch}(x)
&=	
	\frac{1}{x - i \ep}
	+
	\sum_{n = 0}^{\infty}
	b_{n}
	x^{2n+1}
\nn
&=
	i
	\pi
	\delta(x)
	+
	P
	\bigg(
		\frac{1}{x}
		+
		\sum_{n = 0}^{\infty}
		b_{n}
		x^{2n+1}
	\bigg)
\nn
&=
	i
	\pi
	\delta(x)
	+
	P
	\big(
		\text{csch}(x)
	\big)
	\,.
\end{align}
Using this result in \cref{eq:FiniteTempCFTCorr}, we have
\begin{align}
	\corr{0,T}(t,t_{1})
&=
	g^{2}
	\delta(t-t_{1})
	-
	i
	4 
	g^{2} 
	T
	\text{csch}
	\big[
		2
		\pi
		T
        (t-t_{1})
	\big]
\nn
&\hspace{11.5pt}
	-
	\frac{2 g^{2} T^{2}}{L}
	\int
	\dd x
	\dd x_{1}
\nn
&\hspace{26.5pt}
    \times
	P
	\Big(
		\text{csch}
		\big[
			\pi
			T
			\big(
				x_{1}
				-
				x
				+
                (t-t_{1})
			\big)
		\big]
\nn
&\hspace{51.5pt}
        \times
		\text{csch}
		\big[
			\pi
			T
			(
				x_{1}
				-
				x
				-
                (t-t_{1})
			)
		\big]
	\Big)
    \,.
\end{align}
For the remaining integral, we impose that $ x_{1} \neq x \pm (t-t_{1}) $ under the principal value, and use that
\begin{align}
	\text{csch}
	\big(
		\pi 
		T
		x
	\big)
=
	\frac{1}{T}
	\int_{-\infty}^{\infty}
	\frac{\dd k}{2 \pi i}
	\tanh
	\Big[
		\frac{k}{2 T}
	\Big]
	e^{
		-
		i
		k
		x
	}
	\,,
\end{align}
to write
\begin{align}
    \corr{0,T}(t,t_{1})
&=
    2
	g^{2}
	\delta(t-t_{1})
	-
	i
	4 
	g^{2} 
	T
	\text{csch}
	\big[
		2
		\pi
		T
        (t-t_{1})
	\big]
\nn
&\hspace{11.5pt}
	+
    \mathcal{C}_{T}(t,t_{1})
    \,.
\end{align}
where $ \mathcal{C}_{T}(t,t_{1}) $ is given by \cref{eq:CPEnvironmentCorrfiniteT}.
In the infinite temperature limit, the imaginary piece vanishes and the $ \mathcal{C}_{\infty}(t,t_{1}) = - g^{2} \delta(t - t_{1}) $, leaving us with 
\begin{align}
	\corr{0,\infty}(t,t_{1})
&=
	g^{2}
	\delta(t-t_{1})
	\,,
\end{align}
which matches \cref{eq:CriticalCorrInfTemp}.

\section{Exact dynamics from Loschmidt Echo}
\label{app:LoschEcho}

Let $\hat{H}_{\tS \tE}=\hat{H}_{\tS}+\hat{H}_{\tE}+\hat{H}_{\rm int}$ be the Hamiltonian describing a system interacting with an environment. If the interaction is of the form,
\begin{align}
	\hat{H}_{\rm int}
    &=
    \sum_j \hat{P}_j\otimes \hat{V}_j
\end{align}
where $\hat{P}_j$ are the projectors onto the system eigenstates, $\ket{\psi_j}$, then the full Hamiltonian may be expressed as 
\begin{align}
	\hat{H}_{\tS \tE}
    &=
    \sum_j \hat{P}_j\otimes\qty(\epsilon_j+\hat{H}_{\tE} + \hat{V}_j).
\end{align}
Here $\epsilon_j$ are the system eigenenergies and the tensor product has been suppressed. This yields eigenstates of the full Hamiltonian of the form $\ket{\Phi_j^\alpha}=\ket{\psi_j}\otimes\ket{\chi^\alpha_j}$. From this, we evolve
a pure, factorizable initial state, $\ket{\Psi}=(\sum_j c_j\ket{\psi_j})\otimes \ket{\chi_0}$, and find
\begin{align}
	\ket{\Psi(t)}
    &=
    \sum_j e^{-i\epsilon_j t}\ket{\psi_j} \otimes (e^{-i(\hat{H}_{\tE}+\hat{V}_j)t}\ket{\chi_0}).
\end{align}
Tracing out the environment we find the reduced system density matrix
\begin{align}
	\hat{\rho}(t)
    &=
    \sum_{jk} \hat{\rho}_{jk}(0)e^{-i(\epsilon_j-\epsilon_k) t}\ev*{\hat{M}_{jk}(t)}\op{\psi_j}{\psi_k}.
    \label{eq:LoschmidtSolution}
\end{align}
We have defined the echo operator
\begin{align}
	\hat{M}_{jk}(t)
    &=
    e^{i(\hat{H}_{\tE}+\hat{V}_k)t}e^{-i(\hat{H}_{\tE}+\hat{V}_j)t}
\end{align}
and the expectation value is taken in the initial environment state $\ket{\chi_0}$. Taking a derivative in time, the terms vanish and we find the equation of motion
\begin{align}
	\partial_t\hat{\rho}(t)
    =&
    -i[\hat{H}_{\tS},\,\hat{\rho}(t)] 
    \nonumber \\
    &+ \sum_{j\neq k} \hat{\rho}_{jk}(0)e^{-i(\epsilon_j-\epsilon_k) t}\partial_t \ev*{\hat{M}_{jk}(t)}\op{\psi_j}{\psi_k}.
\end{align}
From the solution given in Eq.~(\ref{eq:LoschmidtSolution}), this may be written more explicitly as
\begin{align}
	\partial_t\hat{\rho}(t)
    =&
    -i[\hat{H}_{\tS},\,\hat{\rho}(t)] + \sum_{j\neq k}\frac{\partial_t \ev*{\hat{M}_{jk}(t)}}{\ev*{\hat{M}_{jk}(t)}}\hat{P}_j\hat{\rho}(t)\hat{P}_k.
\end{align}
Using the fact that $\hat{M}_{jk}(t)=\hat{M}_{kj}^\dag(t)$, 
\begin{align}
	\partial_t\hat{\rho}(t)
    =&-i\qty[\hat{H}_{\tS},\,\hat{\rho}(t)]
    \nonumber \\
    &
    +i\sum_{j > k}\Im\qty[\frac{\partial_t \ev*{\hat{M}_{jk}(t)}}{\ev*{\hat{M}_{jk}(t)}}](\hat{P}_j\hat{\rho}(t)\hat{P}_k-\hat{P}_k\hat{\rho}(t)\hat{P}_j)
    \nonumber \\
    &
    + \sum_{j > k}\Re\qty[\frac{\partial_t \ev*{\hat{M}_{jk}(t)}}{\ev*{\hat{M}_{jk}(t)}}](\hat{P}_j\hat{\rho}(t)\hat{P}_k+\hat{P}_k\hat{\rho}(t)\hat{P}_j).
\end{align}
Note that this equation is exact to all orders and may be solved analytically in many cases. In the simplest case of a two-level system, there is only one relevant echo operator, $\hat{M}(t)\equiv\hat{M}_{01}(t)$, and thus the equation of motion simplifies to
\begin{align}
	\partial_t\hat{\rho}(t)
    =&-i[\hat{H}_{\tS},\,\hat{\rho}(t)]
    -i\Omega(t)\qty[\hat{\sigma}_z, \hat{\rho}(t)]
    +\Gamma(t)\mathcal{D}[\hat{\sigma}_z]\hat{\rho}(t),
\end{align}
with Lamb shift and dephasing rates defined as,
\begin{subequations}
\begin{align}
	\Gamma(t)
    &=
    -\frac12 \Re\partial_t \ln \ev*{\hat{M}(t)}
    \\
	\Omega(t)
    &=
    \frac12 \Im\partial_t \ln \ev*{\hat{M}(t)},
\end{align}
\end{subequations}
respectively. The rate may be further simplified in terms of the Loschmidt echo, $L(t)=|\ev*{\hat{M}(t)}|^2$.
\begin{align}
	\Gamma(t)
    &=
    -\frac14 \partial_t \ln L(t)
    \label{eq:LoschmidtgammaApp}
\end{align}
We can now this result to calculate the dephasing rate induced on a qubit due to its coupling to a TFIM using the Loschmidt echo. To this end, we rewrite the full Hamiltonian in \cref{eq:QICHamiltonian} as a tensor sum of qubit projection operators,
\begin{align}
\hH_{\tS \tE} = \hH_{\tS} +  \ket{g} \bra{g} \otimes \hH_{+} + \ket{e} \bra{e} \otimes \hH_{-} 
\end{align}
where
\begin{align}
\label{eq:ICHamiltonianES}
    \hH_{\pm}
&=
	-
    J
    \sum_{j}
    \hsig^{z}_{j}
    \hsig^{z}_{j+1}
    -
    h_{\pm}
    \sum_{j}
    \hsig^{x}_{j},
\end{align}
with $ h_{\pm} =  h \pm \frac{g}{\sqrt{L}} $. Thus. the interaction can be thought of as a system-induced shift in the TFIM magnetic field.

The Loschmidt echo for this system is calculatd as~\cite{Quan2006}
\begin{align}
\label{eq:LoschmidtThetaEp}
	L(t)
&=
	\left|
		\langle
			e^{
				i
				\hH_{+}
				\left(
					t
					-
					t_{0}
				\right)
			}
			e^{
				-
				i
				\hH_{-}
				\left(
					t
					-
					t_{0}
				\right)
			}
		\rangle_{0}
	\right|^{2}
\nn
&=
	\prod_{n = 0}^{N/2-1}
	\left[
		1
		-
		\sin^{2}
		\left(
			\theta_{+,k_{n}}
			-
			\theta_{-,k_{n}}
		\right)
		\sin^{2}
		\left(
			\ep_{-,k_{n}}
			t
		\right)
	\right]
	\,,
\end{align}
where the expectation value is taken in the ground state of the TFIM, with
\begin{align}
\label{eq:ThetaPm}
	\theta_{\pm, k}
&=
	\arctan
	\Big(
		-
		\frac{ 
			r_{k}
		}{
			2 J \cos(a k) - 2h_{\pm}
		}
	\Big)
	\,,
\\
\label{eq:EpPm}
	\ep_{\pm, k} 
&= 	
    2
	\sqrt{
        J^{2}
        +
        h_{\pm}^{2}
        -
        2 J h_{\pm} \cos(a k)
	} 
	\,.
\end{align}
Using \cref{eq:LoschmidtThetaEp} with \cref{eq:ThetaPm,eq:EpPm} in \cref{eq:LoschmidtgammaApp}, and taking the thermodynamic limit, $ L \to \infty $, we find that 
\begin{align}
	\rate{\mu,0}(t)
&=
	\frac{
		2 g^{2}
	}{
		\pi
	}
	\int_{0}^{\frac{\pi}{a}}
	\dd k
    \frac{
        r_{k}^{2}
    }{
        \ep_{k}^{3}
    }
    \sin
    \left(
        2
        \ep_{k}
        t
    \right).
\label{eq:LoschmidtRateapp}
\end{align}
The QFT limit of the equation above is precisely \cref{eq:LoschmidtRate}. 
%
\section{Steady-state emission spectrum}
\label{app:SSEmissionSpectrum}
%
We outline the calculation of the steady-state emission spectrum of the probe, using the frequency-domain prescription detailed in \cite{keefe2025}.
We start from \cref{eq:BornME}, setting $t_0 = 0$ for simplicity, and transform back to the Schr\"odinger picture while maintaining time-nonlocality, to obtain the integro-differential equation
\begin{align}
	\dv{\sdo(t)}{t}
    =
    \sL\sdo(t)+\int_{0}^t \dd{t_1}\fK(t-t_1)\sdo(t_1)
    \,,
\label{eq:TimeNonLocalQME}
\end{align}
where we have introduced the system Liouvillian superoperator, $\sL$, and the memory kernel, $\fK(t)$. The superoperator memory kernel, $\mathcal{K}^{(2)}(t)$, for the time-non-local master equation is written as
\begin{align}
    \mathcal{K}^{(2)}(t)\bullet
    &= \hspace{-2pt}
    -\tr_{E}\qty[\hat{H}_I(0), \qty[\hat{H}_I(-t), e^{i \hat{H}_S t} \bullet e^{-i \hat{H}_S t}\hat{\rho}_E(\beta)]].\nonumber \\
\end{align}
Expressed in matrix form in the usual basis, it is found to be
\begin{align}
    \mathcal{K}^{(2)}(t)
    &=
    -\corr{\mu, T}(0, t)\mqty(
        0  &         0          &          0         & 0  \\
             0           &e^{i\omega_S t}     &          0         &        0           \\
            0            &                   &e^{-i\omega_S t}    &          0         \\
         0&0                   &             0      &0
    ). 
\end{align}
Transforming to the frequency domain via a Fourier-Laplace transform, we have
\begin{align}
    \mathcal{K}^{(2)}[\omega]
    &=
    -\mqty(
        0  &         0          &          0         & 0  \\
             0           &\corr{\mu, T}[\omega-\omega_S]     &          0         &        0           \\
            0            &                   &\corr{\mu, T}[\omega+\omega_S]     &          0         \\
         0&0                   &             0      &0
    ),
\label{eq:Kernel}
\end{align}
where 
\begin{align}
    f[\omega]
    &\equiv
    \int_0^\infty \dd{t} f(t)e^{-i \omega t}.
\end{align}
Note that the quantities in the frequency domain are differentiated from those in the time domain by using square brackets to enclose their arguments. The normalized steady-state emission spectrum is defined by 
\begin{align}
    S[\omega]
    &\equiv
    \frac{1}{2\pi}\int_{-\infty}^\infty \dd{t}  \frac{\ev*{\hsig_+(t)\hsig_-}_{ss}}{\ev*{\hsig_+\hsig_-}_{ss}}e^{-i\omega t},
\end{align}
or, equivalently, in terms of the two-point correlation function in the frequency-domain,
\begin{align}
    S[\omega]
    &\equiv
    \frac{1}{\pi}\Re \frac{\ev*{\hsig_+[\omega]\hsig_-}_{ss}}{\ev*{\hsig_+\hsig_-}_{ss}},
\end{align}
where the expectation value is calculated in the steady-state of the qubit calculated from,
\begin{align}
	\sdo^{ss} \equiv \lim_{\omega \rightarrow 0} \sdo[\omega]
    =
    \lim_{\omega \rightarrow 0}\frac{1}{i\omega-\sL-\fK[\omega]}\sdo(0).
\label{eq:FDQMESoln}
\end{align}
Using the memory kernel in Eq.~(\ref{eq:Kernel}) to then calculate this correlator, we simply find
\begin{align}
    \ev*{\hsig_+[\omega]\hsig_-}_{ss}
    &=
    \frac{\ev*{\hsig_+\hsig_-}_{ss}}{i(\omega-\omega_S)+\corr{\mu, T}[\omega-\omega_S]}.
\end{align}
From this, the normalized spectrum expressed in terms of detuning from the system frequency, $\Delta=\omega-\omega_S$, is found to be
\begin{align}
    S[\Delta]
    &=
    \frac{1}{\pi}\Re\qty[\frac{1}{i\Delta-\corr{\mu, T}[\Delta]}]
    \nonumber \\
    &=
    \frac{1}{\pi}\frac{\Re \corr{\mu, T}[\Delta]}{(\Delta+\Im \corr{\mu, T}[\Delta])^2+(\Re \corr{\mu, T}[\Delta])^2}
    \label{eq:Spec_General}
\end{align}
Importantly, the commutation of the interaction with the qubit free Hamiltonian leads to a spectrum which depends strictly on detuning from the probe frequency $\Delta$ rather than the bare probe frequency $\omega_{S}$ \cite{keefe2025}.
%
\subsection{Zero-temperature}
%
At zero-temperature and finite-gap, the TFIM correlator is given by
\begin{align}
    \corr{\mu, 0}(0, t)
    &=
    4g^2\int_{-\infty}^{\infty} \frac{\dd{k}}{2\pi} \frac{r_k^2}{\epsilon_{k}^2}\cos({2 \epsilon_{k} t}),
\end{align}
in the STF limit. Taking a a Fourier-Laplace transform and applying the Sokhotski–Plemelj theorem, we find
\begin{subequations}
\begin{align}
    \Re \corr{\mu, 0}[\Delta]
    &=
    g^2\hspace{-3pt}\int_{-\infty}^{\infty} \dd{k} \frac{r_{k}^2}{\epsilon_{k}^2} \qty{\delta(\Delta-2\epsilon_{k}) + \delta(\Delta+2\epsilon_{k})}
    \\
    \Im \corr{\mu, 0}[\Delta]
    &=
    -2g^2\mathcal{P}\hspace{-3pt}\int_{-\infty}^{\infty} \frac{\dd{k}}{2\pi} \frac{r_{k}^2}{\epsilon_{k}^2} \qty{\hspace{-2pt}\frac{1}{\Delta-2\epsilon_{k}} + \frac{1}{\Delta+2\epsilon_{k}}\hspace{-2pt}}.
\end{align}
\end{subequations}
For frequencies above the gap, $\Delta^2>4\mu^2$, we find
\begin{align}
\Re \corr{\mu, 0}[\Delta]
    &=
    \frac{g^2\sqrt{\Delta^2-4\mu^2}}{|\Delta|}, 
\end{align}
whereas for all other frequencies the real part is vanishes. Simplifying the imaginary part of the memory kernel, we find
\begin{align}
    \Im \corr{\mu, 0}[\Delta]
    &=
    -8g^2\mathcal{P}\hspace{-3pt}\int_{0}^{\infty} \frac{\dd{k}}{2\pi} \frac{r_{k}^2}{\epsilon_{k}^2}\frac{\Delta}{\Delta^2-4\epsilon_{k}^2}.
\end{align}
Changing the variable of integration to $\epsilon_{k}$, we find a Kramers-Kronig-like form,
\begin{align}
    \Im \corr{\mu, 0}[\Delta]
    &=
    16\mathcal{P}\hspace{-3pt}\int_{|\mu|}^{\infty} \frac{\dd{\epsilon_k}}{2\pi} \Re \corr{\mu,0}[2\epsilon_k]\frac{\Delta}{\Delta^2-4\epsilon_{k}^2}.
\end{align}
Just as in the case of real part, we first consider the frequencies $\Delta^2 > 4\mu^2$. For this regime, the pole lies in the region of integration leading to 
\begin{align}
    \Im \corr{\mu, 0}[\Delta]
    &=
    -2g^2\frac{|\mu|}{\Delta}.
\end{align}
If the pole lies below the gap, $\Delta^2 < 4\mu^2$, we instead find
\begin{align}
    \Im \corr{\mu, 0}[\Delta]
    &=
    -\frac{g^2}{\Delta}\qty(2|\mu|-\sqrt{4\mu^2-\Delta^2}).
\end{align}
Thus, the total emission spectrum may be written as 
\begin{align}
    S[\Delta]
    &=
    \frac{g^2}{\pi}\frac{\sqrt{1-\frac{4\mu^2}{\Delta^2}}}{\Delta^2 + g^4+4g^2|\mu|}\Theta(\Delta^2-4\mu^2).
\end{align}
Notice that for $\Delta^2\gg 4\mu^2$, this reduces to a simple Lorentzian of width roughly given by $g^2/2$. It would appear that the spectrum vanishes whenever $\Delta^2<4\mu^2$, however $\Im \corr{\mu, 0}[0]=0$, and so the expression in Eq.~\ref{eq:Spec_General} appears to diverge. Analytically continuing and again applying the Sokhotski–Plemelj theorem, we find
\begin{align}
    \lim_{\epsilon\to 0^+}\frac{1}{i(\Delta - \Im \corr{\mu, 0}[\Delta]) + \epsilon}
    =&
    \pi\delta(\Delta-\Im \corr{\mu, 0}[\Delta]) 
    \nonumber \\
    &
    -i \mathcal{P}\qty(\frac{1}{\Delta-\Im \corr{\mu, 0}[\Delta]}).
\end{align}
Since the emission spectrum is proportional to the real part, we find an additional delta function contribution at the qubit frequency. This arises as a direct result of the vanishing decay rate leading to fractional dephasing in the steady-state. Thus, we have the final expression
\begin{align}
    S[\Delta]
    &=
    \Phi_0\delta(\Delta)
    +
     \frac{g^2}{\pi}\frac{\sqrt{1-\frac{4\mu^2}{\Delta^2}}}{\Delta^2 + g^4+4g^2|\mu|}\Theta(\Delta^2-4\mu^2),
\end{align}
where we have rewritten the delta function in terms of 
\begin{align}
    \Phi_0
    &=
    \frac{4|\mu|}{4|\mu|+g^2}.
\end{align}
Notice that in the limit of criticality, $\mu\to 0$, we have $\corr{0, 0}[\Delta]=-g^2$, and we recover a fully Lorenztian emission spectrum,
\begin{align}
    S[\Delta]
    &=
    \frac{1}{\pi}\frac{\rate{0, 0}}{\Delta^2 + \rate{0, 0}^2},
\end{align}
which is a characteristic of Markovian dynamics.

\subsection{Finite-temperature}

From \cref{eq:RealPartCorr}, the correlator at finite-temperature and zero-gap reduces to 
\begin{align}
	\Re \corr{0, T}(0, t)
    =&
    \Re\corr{0,0}(0, t)
    +
    \mathcal{C}_T(0, t).
\end{align}
We have already shown $\corr{0,0}[\Delta]=\rate{0,0}$, and so we will now calculate $\mathcal{C}_{T}[\Delta]$. Unlike in the zero temperature case, we integrate over $k$ before moving to the frequency domain. From \cref{eq:CPEnvironmentCorrfiniteT}, we have
\begin{align}
	\mathcal{C}_{T}(0, t)
    =&
    -8g^2 t T^2\csch(2\pi t T).
\end{align}
Now taking the Fourier-Laplace transform, we find
\begin{align}
	\mathcal{C}_{T}[\Delta]
    =&
    -\frac{g^2}{\pi^2}\zeta\qty(2, \frac{1}{2} + i\frac{\Delta}{4\pi T}),
\end{align}
where $\zeta$ is the generalized Riemann zeta function. While the imaginary part has no closed form, the real part may be expressed simply as
\begin{align}
	\Re \mathcal{C}_{T}[\Delta]
    =&
    -\frac{g^2}{2}\sech^2\qty(\frac{\Delta}{4T}).
\end{align}
From this, we write the emission spectrum as
\begin{align}
	S[\Delta]
    =&
    \frac{1}{\pi} \frac{\rate{0, 0}\qty(1-\frac{1}{2}\sech^2\qty(\frac{\Delta}{4T}))}{(\Delta-\Im \mathcal{C}_{T}[\Delta])^2+\rate{0, 0}^2\qty(1-\frac{1}{2}\sech^2\qty(\frac{\Delta}{4T}))^2}.
\end{align}
In the limit of both small and large temperatures, imaginary part of $\mathcal{C}_{T}[\Delta]\to0$ and the the secant term becomes constant, resulting in the expected Lorentzian spectra of widths $\rate{0,0}$ and $\rate{0, 0}/2=\rate{0, \infty}$, respectively. 

\section{Quantifying non-Markovianity}
\label{app:Quantifying non-Markovianity}
%
\begin{figure}[b!]
\begin{center}
    \includegraphics[width=\columnwidth]{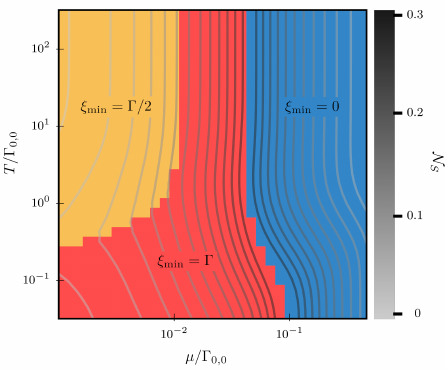}
    \caption{Plot showing $\mathcal{N}_S$ (Contours) using only the spectra at the fixed points as the Markovian ansatz and the corresponding $\xi$ (Shading).  
 }
\label{fig:FiniteTemperature}
\end{center}
\end{figure}
%
In order to quantify non-Markovianity, we employ a measure similar to the spectral measure proposed in \cite{keefe2025},
\begin{align}
	\mathcal{N}_S
    &=
    \frac{1}{\Omega}D( S || S_M) 
    \nonumber \\ 
    &= \frac{1}{\Omega}\int_{-\infty}^\infty \dd{\omega} S[\omega]\log_2 \frac{S[\omega]}{S_{M}[\omega]},
\end{align}
which uses the Kullback-Leibler divergence (KDL) between the non-Markovian spectrum and some Markovian limit, $S_M[\omega]$, to quantify deviation from Markovian dynamics. In our system, however, the KLD is not suitable as the vanishing spectrum within the gap leads to a diverging cross entropy. Rather, we consider the Jensen-Shannon divergence (JSD),
\begin{align}
	\jsd(S|| S_M)
    &=
    \frac12\qty(D(S|| \bar{S}) + D(S_M || \bar{S})),
\end{align}
where we have defined
\begin{align}
	\bar{S}[\omega]
    &\equiv
    \frac12\qty(S[\omega + S_M[\omega]).
\end{align}
%
%
\begin{figure*}[t!]
\begin{center}
    \includegraphics[width=236pt]{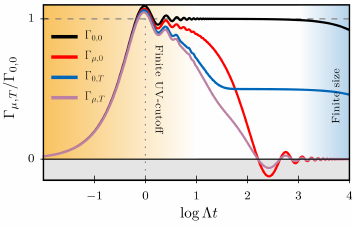}
    \includegraphics[width=\columnwidth]{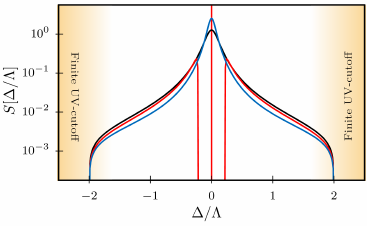}
    \caption{(Left) Induced time-dependent rates for different regimes in the STF limit. All are calculated using the discrete sum given in \cref{eq:RateFiniteUV} with $N=10^6$. Each rate resembles the corresponding rate in the QFT limit, however all of them show a similar behavior not seen previously. This is due to the finite energy scale given by the UV-cutoff, $\Lambda$. At late times, the rates diverge from the QFT limit only due to finite-size effects. (Right) Steady-state emission spectra in the SFT limit calculated analytically in the using eqs.~(\ref{eq:SFTCorrFDReal}-\ref{eq:SFTCorrFDRealTherm}). Here there is no finite size effects, however the effect of the UV-cutoff manifests at large detunings. The finite-gap, finite temperature case is omitted both for clarity and due to the lack of analytics for $\Im G_{\mu, T}[\Delta]$. 
 }
\label{fig:FiniteUVCutoff}
\end{center}
\end{figure*}
%
Another consideration is appropriate choice for the reference Markovian spectrum in the case of the TFIM, where the probe spectrum is exactly Markovian in three separate limits, (i) at the critical point, (ii) at zero gap, infinite temperature, and (iii) at infinite gap. At all these points, the spectrum becomes a Lorentzian, but each with a distinct width. We describe this variation by considering a Lorentzian ansatz,
\begin{align}
	S_M[\omega] = S_\xi[\omega]
    &=
    \frac{1}{\pi}\frac{\xi}{\omega^2+\xi^2}.
\end{align}
and defining a modified spectral measure 
\begin{align}
	\mathcal{N}_S
    &=
    \min_{\xi}\qty{ \jsd(S||S_\xi)}.
\end{align}
where the reference Markovian spectrum width $\xi$ at an arbitrary point in $\qty{\mu, T}$ phase diagram is constrained by minimizing the JSD at that point. Throughout this work, we consider both minimization over a general width, $\xi\in [0, \infty)$, as well as over the widths set by the spectra at the fixed points, $\xi\in\qty{\rate{0,0}, \rate{0, \infty}, 0}$.
%
\section{Statistical field theory limit}
\label{app:StatisticalFieldTheoryLimit}
%
Here we present the effect of a finite UV-cutoff set by a non-zero value of the lattice constant $a$, on the non-Markovian rates and spectra obtained for the probe qubit. Beginning from \cref{eq:CorrDisc} and evaluating the four-point expectation values, we find
\begin{align}
	\corr{\mu,T}(t,t_{1})
&=
    \frac{4 g^{2}}{L}
    \sum_{m=-N/2}^{N/2-1}
	\frac{
		r_{k_m}^{2}
	}{
		\ep_{k_m}^{2}
	}
	\frac{
		\cos
		[
			2
			\ep_{k_m}
			(
				t
				-
				t_{1}
				+
				\frac{
                    i
				}{
					2
                    T
				}
			)
		]
	}{
		1
		+
		\cosh
		\left[
			\ep_{k_m}/T
		\right]
	}
	\,.
\end{align}
While previously this was converted to an integral, as in \cref{eq:FiniteEnvironmentCorr}, analytic expression remain intractable without the QFT limit given in \cref{eq:ContinuumSubs}. Instead, the discrete sum is retained and the rate is calculated using \cref{eq:gamma},
\begin{align}
\label{eq:RateFiniteUV}
	\rate{\mu,T}(t)
&=
    \frac{2 g^{2}}{L}
    \sum_{m=-N/2}^{N/2-1}
	\frac{
		r_{k_m}^{2}
	}{
		\ep_{k_m}^{3}
	}
	\frac{
		\sin
		[
			2
			\ep_{k_m}
			(
				t
				+
				\frac{
                    i
				}{
					2
                    T
				}
			)
		]
	}{
		1
		+
		\cosh
		\left[
			\ep_{k_m}/T
		\right]
	}
	\,.
\end{align}
This is a discrete quantity and thus is subject to finite size effects. Since the timescale at which these effects manifest scales linearly with $N$, a value of $N=10^6$ is more than sufficient to show the short-time UV effects. As shown in \cref{fig:FiniteUVCutoff}, all regimes show a fast ramp to a slightly higher value than $\rate{0, 0}$, followed by damped oscillations about the corresponding rate in the QFT limit. Naturally, all four curves should match the QFT limit as $t\to\infty$, however the finite size effects begin to dominate at later times.

The transient peak in $\Gamma_{\mu,T}(t)$, which is nearly independent of $\qty{\mu,T}$, is an attribute of the finite UV cutoff. We can analytically estimate it at the QCP, which is also where seeing UV effects is most counter-intuitive. From \cref{eq:FiniteEnvironmentCorr},
\begin{equation}
\begin{aligned}
\Gamma_{0,0}^{\text{SFT}}(t) &= \frac{2g^2}{\pi} \int_0^{\pi/a} \mathrm{d}k \, \frac{r_k^2}{\epsilon_k^3} \sin(2\epsilon_k t) \\
&= \Gamma_{0,0}\left(\frac{2}{\pi} \int_0^{\pi/2} \mathrm{d}x \, \frac{\cos^2 x}{\sin x} \sin(y \sin x)\right),
\end{aligned}
\label{eq:GamSFTQCP}
\end{equation}
where the variables $y = 4\Lambda_{} t,~\Lambda_{} = 2J,~x = ak/2$, $\epsilon_k= 2\Lambda \sin(ak/2), r_k = 2J \sin(ak) = \epsilon_k \cos(ak/2)$. Using  ${v=a\Lambda=1}$ in the QFT limit, leads to $\Gamma_{0,0} = g^2$. The integral is evaluated using the Bessel integral identity,
\begin{align}
&\frac{\mathrm{d}}{\mathrm{d}y} \left[ \frac{\Gamma_{0,0}^{\text{SFT}}(t)}{\Gamma_{0,0}} \right] = \frac{2}{\pi} \int_0^{\pi/2} \mathrm{d}x \, \cos^2 x \cos(y \sin x) = \frac{J_1(y)}{y} \nonumber\\
&\implies \frac{\Gamma_{0,0}^{\text{SFT}}(t)}{\Gamma_{0,0}} = \int_0^y \frac{J_1(u)}{u} \, \mathrm{d}u.
\end{align}
Then the extremum condition of $\Gamma_{0,0}^{\text{SFT}}(t)$ begets the first peak set by $J_1(y) = 0$, in the left plot of \cref{fig:FiniteUVCutoff}, at
\begin{equation}
\begin{aligned}
t_{\text{peak}} \approx \frac{0.96}{\Lambda},~\left( \frac{\Gamma_{0,0}^{\text{SFT}}}{\Gamma_{0,0}} \right)_{\text{max}} \approx 1.1 .
\end{aligned}
\end{equation}
In the long-time limit, $\Gamma_{0,0}^{\text{SFT}}(t\to\infty) \to \Gamma_{0,0}$ as per the QFT limit. The physical mechanism behind this rate enhancement at short times becomes clear by reformulating \cref{eq:GamSFTQCP} in the spectral domain as,
\begin{align}
&\Gamma_{0,0}^{\text{SFT}}(t) = \int_0^{2\Lambda} \mathrm{d}\epsilon_k ~J(\epsilon_k)\, \frac{\sin(2\epsilon_k t)}{\epsilon_k}.
\end{align}
which is the lattice counterpart of \cref{eq:FiniteEnvironmentRatevac} at the QCP, with
\begin{align}
J(\omega) \equiv  \frac{2g^2}{\pi} \sqrt{1 - \frac{\omega^2}{16J^2}}\Theta(\omega-4J).
\end{align}
In the QFT-limit, a flat spectral density $J_{\text{QFT}}(\omega) = \frac{2g^2}{\pi}$ yields $\Gamma_{0,0}(t)=g^2$. On the other hand, SFT shows varying interference effects at different time-scales: \newline
(i) For $t\ll 1/(2\Lambda)$, all TFIM modes with energy bounded by $2\Lambda$ oscillate in-phase ($\sin(\omega t) > 0$) increasing up to $2\Lambda t_{\text{peak}}\sim 1$. The precise location of $t_{\text{peak}}$ around this order of magnitude estimate is governed by the band-edge regularized form of $J(\omega)$. \newline
(ii) For $t \gtrsim 1/2\Lambda$, the contributions from modes near TFIM the band edge start interfering destructively.
\par
We take a different approach for calculating the emission spectrum. Since we consider a Fourier-Laplace transform over the correlator, some analytics can be obtained in the continuum limit. Transforming \cref{eq:FiniteEnvironmentCorr} to the frequency domain [see \cref{app:SSEmissionSpectrum}], we find,
\begin{subequations}
\begin{align}
    \Re \corr{\mu, T}[\Delta]
    &=
    -g^2\hspace{-3pt}\int_{0}^{\frac{\pi}{a}} \dd{k} \frac{r_{k}^2}{\epsilon_{k}^2} \frac{\cosh\qty[\epsilon_k/T]}{\cosh^2[\epsilon_k/2T]}
    \nn
    &\hspace{45pt}
    \times\qty{\delta(\Delta-2\epsilon_{k}) + \delta(\Delta+2\epsilon_{k})},
    \\
    \Im \corr{\mu, T}[\Delta]
    &=
    -g^2\mathcal{P}\hspace{-3pt}\int_{0}^{\frac{\pi}{a}} \frac{\dd{k}}{\pi} \frac{r_{k}^2}{\epsilon_{k}^2}\frac{\cosh\qty[\epsilon_k/T]}{\cosh^2[\epsilon_k/2T]}
    \nn
    &\hspace{45pt}
    \times\qty{\hspace{-2pt}\frac{1}{\Delta-2\epsilon_{k}} + \frac{1}{\Delta+2\epsilon_{k}}\hspace{-2pt}}.
\end{align}
\end{subequations}
This may be further split into temperature-dependent and -independent contributions, $\corr{\mu, T}=\corr{\mu, 0} + G_{\mu, T}$, where 
\begin{subequations}
\begin{align}
    \Re \corr{\mu, 0}[\Delta]
    &=
    2g^2\hspace{-3pt}\int_{0}^{\frac{\pi}{a}} \dd{k} \frac{r_{k}^2}{\epsilon_{k}^2} \qty{\delta(\Delta-2\epsilon_{k}) + \delta(\Delta+2\epsilon_{k})},
    \\
    \Im \corr{\mu, 0}[\Delta]
    &=
    -2g^2\mathcal{P}\hspace{-3pt}\int_{0}^{\frac{\pi}{a}} \frac{\dd{k}}{\pi} \frac{r_{k}^2}{\epsilon_{k}^2}\qty{\hspace{-2pt}\frac{1}{\Delta-2\epsilon_{k}} + \frac{1}{\Delta+2\epsilon_{k}}\hspace{-2pt}}, 
\end{align}
\end{subequations}
corresponding to the full correlator at zero-temperature, and 
\begin{subequations}
\begin{align}
    \Re G_{\mu, T}[\Delta]
    &=
    -g^2\hspace{-3pt}\int_{0}^{\frac{\pi}{a}} \dd{k} \frac{r_{k}^2}{\epsilon_{k}^2} \sech^2[\epsilon_k/2T]
    \nn
    &\hspace{45pt}
    \times\qty{\delta(\Delta-2\epsilon_{k}) + \delta(\Delta+2\epsilon_{k})},
    \\
    \Im G_{\mu, T}[\Delta]
    &=
    g^2\mathcal{P}\hspace{-3pt}\int_{0}^{\frac{\pi}{a}} \frac{\dd{k}}{\pi} \frac{r_{k}^2}{\epsilon_{k}^2}\sech^2[\epsilon_k/2T]
    \nn
    &\hspace{45pt}
    \times\qty{\hspace{-2pt}\frac{1}{\Delta-2\epsilon_{k}} + \frac{1}{\Delta+2\epsilon_{k}}\hspace{-2pt}},
\end{align}
\end{subequations}
corresponding to the finite-temperature correction. Considering only the zero-temperature limit, and using the full dispersion relation given in \cref{eq:DispersionBare}, we find for $2|J-h|\le \Delta \le 2|J+h|$,
\begin{subequations}
\begin{align}
    \Re \corr{\mu, 0}[\Delta]
    &=
    \frac{g^2}{a|\Delta|h^2}\sqrt{(2hJ)^2-(J^2+h^2-\Delta^2/4)^2},
    \label{eq:SFTCorrFDReal} \\
    \Im \corr{\mu, 0}[\Delta]
    &=
    -\frac{g^2}{a\Delta h^2}\qty(|J^2-h^2|-\Delta^2/4).
    \label{eq:SFTCorrFDIm}
\end{align}
\end{subequations}
For all other values of $\Delta$, the correlator is strictly imaginary. 

The real part of the temperature dependent contribution may be solve analytically in the same way, 
\begin{align}
    \Re G_{\mu, T}[\Delta]
    &=
    -\frac12 \corr{\mu, T}[\Delta] \sech^2\frac{\Delta}{4 T}.
    \label{eq:SFTCorrFDRealTherm}
\end{align}
The imaginary part, however, cannot be expressed analytically, and so is integrated numerically to calculate the qubit emission spectrum. As shown in \cref{fig:FiniteUVCutoff}, the spectra predictably match the QFT limit in the small frequency regime, including delta-function and TFIM spectral gap. At large frequencies, however, the spectra begin to deviate from the QFT limit, vanishing completely past the UV cutoff, $\lambda=|J+h|$. As expected from the short-time behavior or the rates, this would naturally lead to a small non-zero value $\mathcal{N}_S$, even at the critical point due to the band edges imposed by a finite UV cutoff. 
%
\section{Flow analysis}
\label{app:FlowAnalysis}
%
For any point in the phase diagram of the TFIM environment, $\mathbf{v} = (\mu, T) $, we denote the associated qubit emission spectrum as $ S(\mathbf{v}) $, and define a vector of the spectral distances from each of the fixed points, $ \mathbf{v}^{*}_{1} = (0,0), \mathbf{v}^{*}_{2} = (0,\infty), \mathbf{v}^{*}_{3} = (\infty,0) $, as
\begin{align}
	\mathcal{D}(\mathbf{v})
=
	\left(
		d_{S}(\mathbf{v}, \mathbf{v}^{*}_{1})
		,
		d_{S}(\mathbf{v}, \mathbf{v}^{*}_{2})
		,
		d_{S}(\mathbf{v}, \mathbf{v}^{*}_{3})
	\right).
\end{align}
Here $ d(\mathbf{v}_1, \mathbf{v}_2) = \sqrt{\jsd(S(\mathbf{v}_1)||S(\mathbf{v}_2))} $ with JSD denoting Jensen-Shannon divergence. We can obtain the induced flow in spectral fixed-
point space by taking the directional derivative of $ \mathcal{D}(\mathbf{v}) $, with respect to a vector $\mathbf{u}$ in the $(\mu, T)$ plane, 
\begin{align}
	\mathcal{F}_{\mathbf{u}}(\mathbf{v})
=
	\left(
		\mathbf{u}
		\cdot
		\nabla
	\right)
	\mathcal{D}(\mathbf{v})
	\,.
\end{align}
To probe the flow evenly across scales, we use the logarithmic gradient in the rate-normalized phase space coordinates, $ \nabla = (\partial/ \partial\ln (\mu / \Gamma_{0,0}), \partial/ \partial \ln (T / \Gamma_{0,0})) $.

To visualize the flow in two dimensions, we use a basis of principal components from the distance vectors for each of the fixed points. We start by constructing a $3 \times 3 $ matrix from the distance vectors for each of the fixed points, as
\begin{align}
	\mathcal{B}
=
	\begin{bmatrix}
	\begin{array}{c|c|c}
		&
		&
	 	\\
			\mathcal{D}(\mathbf{v}_{1}^{*}) 
			- 
			\bar{\mathcal{D}}
		&
			\mathcal{D}(\mathbf{v}_{2}^{*})
			- 
			\bar{\mathcal{D}}
		&
			\mathcal{D}(\mathbf{v}_{3}^{*})
			- 
			\bar{\mathcal{D}}
	 	\\
		&
		&
	\end{array}
	\end{bmatrix}
\end{align}
where we have also shifted each vector by $ \bar{\mathcal{D}} = \frac{1}{3} \sum_{i} \mathcal{D}(\mathbf{v}_{i}^{*}) $ so that the origin is at the center of mass of the points. We then perform a singular value decomposition on $ \mathcal{B} $ writing
\begin{align}
	\mathcal{B}
&=
	\mathcal{V}
	\Sigma\hspace{1pt}
	\mathcal{U}^{T}
	\,,
\end{align}
where $ \mathcal{V} \Sigma $ is the principal component score matrix, $ \Sigma $ itself is the diagonal matrix of the ordered singular values of $ \mathcal{B} $, and the columns of $ \mathcal{U} $ are the principal component vectors. Additionally, $ \mathcal{U} $ and $ \mathcal{V} $ are both guaranteed to be orthogonal matrices because $ \mathcal{B} $ is real. 

Since larger singular values correspond to directions of greater variance of the centered fixed points, it is natural to construct a reduced description by projecting onto the principal component vectors that capture the majority of this variance. Here, the third singular value is negligible compared with the first two, so we choose the first two principal component vectors to define a two-dimensional representation of fixed point space, which we denote $ \mathcal{U}_{2} $. We can now map any vector in fixed point space to the two-dimensional representation through the transformation
\begin{align}
	\mathcal{D}_{2}(\mathbf{v})
=
	\mathcal{U}_{2}^{T}
	\left(
		\mathcal{D}(\mathbf{v})
		- 
		\bar{\mathcal{D}}
	\right)
	\,,
\end{align}
and the flow becomes
\begin{align}
\label{eq:2DFlowField}
	\mathcal{F}_{\mathbf{u},2}(\mathbf{v})
=
	\left(
		\mathbf{u}
		\cdot
		\nabla
	\right)
	\mathcal{D}_{2}(\mathbf{v})
	\,.
\end{align}
We visualize \cref{eq:2DFlowField} for the simplest symmetric choice of off-axis direction $ \mathbf{u} = (1,1)/\sqrt{2}$ in \cref{fig:FlowPlot} of the main text.  
%
\subsection{RG beta functions}
\label{sec:RG_beta}
%
Here we briefly derive the conventional RG flow equations of the continuum field theory \cref{eq:S_EMajorana} whose corresponding JSD based flow is shown in \cref{fig:FlowPlot}. A scale transformation $x\to x e^{-\ell}$, yields $z \to z' = z e^{-\ell}$, $\bar{z} \to \bar{z}' = \bar{z} e^{-\ell}$. This informs the rescaling of the field as $\psi \to e^{-\ell/2} \psi'$ in order to keep the kinetic term in the action invariant. Consequently, we obtain the scaling transformation of the mass term as,
\begin{equation}
S_{\text{mass}} = \int (e^{2\ell} d^2z') \mu (e^{-\ell} \bar{\psi}' \psi') = \int d^2z' (\mu e^{\ell}) \bar{\psi}' \psi'.
\end{equation}
Since the theory is non-interacting, this tree level result is exact, and is independent of the temperature which decides the finite size of the imaginary time ($y$) direction --- thus, it does not appear in the Lagrangian density. From the transformed mass $\mu\to \mu_0 e^{\ell}$, we find $\beta_{\mu}=\mu$. 

On the other hand, since both the space and (imaginary) time scales are coarsened by $e^{-\ell}$, the finite size along the $y$ axis in particular also shrinks as,
\begin{equation}
 \beta(\ell) = \beta_0 e^{-\ell}.
\end{equation}
This implies $T(\ell)=T_0e^{\ell}$, or, $\beta_T=T$, which is again independent of $\mu$, and is set entirely by the scaling transformation property of the space-time cylinder under the exact tree level consideration. The identical scaling behavior of $\mu$ and $T$ also follows from the isotropy of underlying Dirac theory in the Euclidean space-time, carrying a dynamic exponent $z=1$ (chapter-5 of \cite{Sachdev2011}).


\begin{thebibliography}{45}%
	\makeatletter
	\providecommand \@ifxundefined [1]{%
		\@ifx{#1\undefined}
	}%
	\providecommand \@ifnum [1]{%
		\ifnum #1\expandafter \@firstoftwo
		\else \expandafter \@secondoftwo
		\fi
	}%
	\providecommand \@ifx [1]{%
		\ifx #1\expandafter \@firstoftwo
		\else \expandafter \@secondoftwo
		\fi
	}%
	\providecommand \natexlab [1]{#1}%
	\providecommand \enquote  [1]{``#1''}%
	\providecommand \bibnamefont  [1]{#1}%
	\providecommand \bibfnamefont [1]{#1}%
	\providecommand \citenamefont [1]{#1}%
	\providecommand \href@noop [0]{\@secondoftwo}%
	\providecommand \href [0]{\begingroup \@sanitize@url \@href}%
	\providecommand \@href[1]{\@@startlink{#1}\@@href}%
	\providecommand \@@href[1]{\endgroup#1\@@endlink}%
	\providecommand \@sanitize@url [0]{\catcode `\\12\catcode `\$12\catcode `\&12\catcode `\#12\catcode `\^12\catcode `\_12\catcode `\%12\relax}%
	\providecommand \@@startlink[1]{}%
	\providecommand \@@endlink[0]{}%
	\providecommand \url  [0]{\begingroup\@sanitize@url \@url }%
	\providecommand \@url [1]{\endgroup\@href {#1}{\urlprefix }}%
	\providecommand \urlprefix  [0]{URL }%
	\providecommand \Eprint [0]{\href }%
	\providecommand \doibase [0]{https://doi.org/}%
	\providecommand \selectlanguage [0]{\@gobble}%
	\providecommand \bibinfo  [0]{\@secondoftwo}%
	\providecommand \bibfield  [0]{\@secondoftwo}%
	\providecommand \translation [1]{[#1]}%
	\providecommand \BibitemOpen [0]{}%
	\providecommand \bibitemStop [0]{}%
	\providecommand \bibitemNoStop [0]{.\EOS\space}%
	\providecommand \EOS [0]{\spacefactor3000\relax}%
	\providecommand \BibitemShut  [1]{\csname bibitem#1\endcsname}%
	\let\auto@bib@innerbib\@empty
	\bibitem [{\citenamefont {Fauseweh}(2024)}]{Fauseweh2024}%
	\BibitemOpen
	\bibfield  {author} {\bibinfo {author} {\bibfnamefont {B.}~\bibnamefont {Fauseweh}},\ }\bibfield  {title} {\bibinfo {title} {Quantum many-body simulations on digital quantum computers: State-of-the-art and future challenges},\ }\href {https://doi.org/10.1038/s41467-024-46402-9} {\bibfield  {journal} {\bibinfo  {journal} {Nature Communications}\ }\textbf {\bibinfo {volume} {15}},\ \bibinfo {pages} {2123} (\bibinfo {year} {2024})}\BibitemShut {NoStop}%
	\bibitem [{\citenamefont {Montenegro}\ \emph {et~al.}(2025)\citenamefont {Montenegro}, \citenamefont {Mukhopadhyay}, \citenamefont {Yousefjani}, \citenamefont {Sarkar}, \citenamefont {Mishra}, \citenamefont {Paris},\ and\ \citenamefont {Bayat}}]{Montenegro2025}%
	\BibitemOpen
	\bibfield  {author} {\bibinfo {author} {\bibfnamefont {V.}~\bibnamefont {Montenegro}}, \bibinfo {author} {\bibfnamefont {C.}~\bibnamefont {Mukhopadhyay}}, \bibinfo {author} {\bibfnamefont {R.}~\bibnamefont {Yousefjani}}, \bibinfo {author} {\bibfnamefont {S.}~\bibnamefont {Sarkar}}, \bibinfo {author} {\bibfnamefont {U.}~\bibnamefont {Mishra}}, \bibinfo {author} {\bibfnamefont {M.~G.}\ \bibnamefont {Paris}},\ and\ \bibinfo {author} {\bibfnamefont {A.}~\bibnamefont {Bayat}},\ }\bibfield  {title} {\bibinfo {title} {Review: Quantum metrology and sensing with many-body systems},\ }\href {https://doi.org/https://doi.org/10.1016/j.physrep.2025.05.005} {\bibfield  {journal} {\bibinfo  {journal} {Physics Reports}\ }\textbf {\bibinfo {volume} {1134}},\ \bibinfo {pages} {1} (\bibinfo {year} {2025})}\BibitemShut {NoStop}%
	\bibitem [{\citenamefont {Khemani}\ \emph {et~al.}(2016)\citenamefont {Khemani}, \citenamefont {Lazarides}, \citenamefont {Moessner},\ and\ \citenamefont {Sondhi}}]{Khemani2016}%
	\BibitemOpen
	\bibfield  {author} {\bibinfo {author} {\bibfnamefont {V.}~\bibnamefont {Khemani}}, \bibinfo {author} {\bibfnamefont {A.}~\bibnamefont {Lazarides}}, \bibinfo {author} {\bibfnamefont {R.}~\bibnamefont {Moessner}},\ and\ \bibinfo {author} {\bibfnamefont {S.~L.}\ \bibnamefont {Sondhi}},\ }\bibfield  {title} {\bibinfo {title} {Phase structure of driven quantum systems},\ }\href {https://doi.org/10.1103/PhysRevLett.116.250401} {\bibfield  {journal} {\bibinfo  {journal} {Phys. Rev. Lett.}\ }\textbf {\bibinfo {volume} {116}},\ \bibinfo {pages} {250401} (\bibinfo {year} {2016})}\BibitemShut {NoStop}%
	\bibitem [{\citenamefont {Skinner}\ \emph {et~al.}(2019)\citenamefont {Skinner}, \citenamefont {Ruhman},\ and\ \citenamefont {Nahum}}]{Swingle2019}%
	\BibitemOpen
	\bibfield  {author} {\bibinfo {author} {\bibfnamefont {B.}~\bibnamefont {Skinner}}, \bibinfo {author} {\bibfnamefont {J.}~\bibnamefont {Ruhman}},\ and\ \bibinfo {author} {\bibfnamefont {A.}~\bibnamefont {Nahum}},\ }\bibfield  {title} {\bibinfo {title} {Measurement-induced phase transitions in the dynamics of entanglement},\ }\bibfield  {journal} {\bibinfo  {journal} {Physical Review X}\ }\textbf {\bibinfo {volume} {9}},\ \href {https://doi.org/10.1103/physrevx.9.031009} {10.1103/physrevx.9.031009} (\bibinfo {year} {2019})\BibitemShut {NoStop}%
	\bibitem [{\citenamefont {Fazio}\ \emph {et~al.}(2025)\citenamefont {Fazio}, \citenamefont {Keeling}, \citenamefont {Mazza},\ and\ \citenamefont {Schirò}}]{Fazio2025}%
	\BibitemOpen
	\bibfield  {author} {\bibinfo {author} {\bibfnamefont {R.}~\bibnamefont {Fazio}}, \bibinfo {author} {\bibfnamefont {J.}~\bibnamefont {Keeling}}, \bibinfo {author} {\bibfnamefont {L.}~\bibnamefont {Mazza}},\ and\ \bibinfo {author} {\bibfnamefont {M.}~\bibnamefont {Schirò}},\ }\bibfield  {title} {\bibinfo {title} {Many-body open quantum systems},\ }\bibfield  {journal} {\bibinfo  {journal} {SciPost Physics Lecture Notes}\ }\href {https://doi.org/10.21468/scipostphyslectnotes.99} {10.21468/scipostphyslectnotes.99} (\bibinfo {year} {2025})\BibitemShut {NoStop}%
	\bibitem [{\citenamefont {Barreiro}\ \emph {et~al.}(2011)\citenamefont {Barreiro}, \citenamefont {M{\"u}ller}, \citenamefont {Schindler}, \citenamefont {Nigg}, \citenamefont {Monz}, \citenamefont {Chwalla}, \citenamefont {Hennrich}, \citenamefont {Roos}, \citenamefont {Zoller},\ and\ \citenamefont {Blatt}}]{Barreiro2011}%
	\BibitemOpen
	\bibfield  {author} {\bibinfo {author} {\bibfnamefont {J.~T.}\ \bibnamefont {Barreiro}}, \bibinfo {author} {\bibfnamefont {M.}~\bibnamefont {M{\"u}ller}}, \bibinfo {author} {\bibfnamefont {P.}~\bibnamefont {Schindler}}, \bibinfo {author} {\bibfnamefont {D.}~\bibnamefont {Nigg}}, \bibinfo {author} {\bibfnamefont {T.}~\bibnamefont {Monz}}, \bibinfo {author} {\bibfnamefont {M.}~\bibnamefont {Chwalla}}, \bibinfo {author} {\bibfnamefont {M.}~\bibnamefont {Hennrich}}, \bibinfo {author} {\bibfnamefont {C.~F.}\ \bibnamefont {Roos}}, \bibinfo {author} {\bibfnamefont {P.}~\bibnamefont {Zoller}},\ and\ \bibinfo {author} {\bibfnamefont {R.}~\bibnamefont {Blatt}},\ }\bibfield  {title} {\bibinfo {title} {An open-system quantum simulator with trapped ions},\ }\href {https://doi.org/10.1038/nature09801} {\bibfield  {journal} {\bibinfo  {journal} {Nature}\ }\textbf {\bibinfo {volume} {470}},\ \bibinfo {pages} {486} (\bibinfo {year} {2011})}\BibitemShut {NoStop}%
	\bibitem [{\citenamefont {Choi}\ \emph {et~al.}(2017)\citenamefont {Choi}, \citenamefont {Choi}, \citenamefont {Landig}, \citenamefont {Kucsko}, \citenamefont {Zhou}, \citenamefont {Isoya}, \citenamefont {Jelezko}, \citenamefont {Onoda}, \citenamefont {Sumiya}, \citenamefont {Khemani}, \citenamefont {von Keyserlingk}, \citenamefont {Yao}, \citenamefont {Demler},\ and\ \citenamefont {Lukin}}]{Choi2017}%
	\BibitemOpen
	\bibfield  {author} {\bibinfo {author} {\bibfnamefont {S.}~\bibnamefont {Choi}}, \bibinfo {author} {\bibfnamefont {J.}~\bibnamefont {Choi}}, \bibinfo {author} {\bibfnamefont {R.}~\bibnamefont {Landig}}, \bibinfo {author} {\bibfnamefont {G.}~\bibnamefont {Kucsko}}, \bibinfo {author} {\bibfnamefont {H.}~\bibnamefont {Zhou}}, \bibinfo {author} {\bibfnamefont {J.}~\bibnamefont {Isoya}}, \bibinfo {author} {\bibfnamefont {F.}~\bibnamefont {Jelezko}}, \bibinfo {author} {\bibfnamefont {S.}~\bibnamefont {Onoda}}, \bibinfo {author} {\bibfnamefont {H.}~\bibnamefont {Sumiya}}, \bibinfo {author} {\bibfnamefont {V.}~\bibnamefont {Khemani}}, \bibinfo {author} {\bibfnamefont {C.}~\bibnamefont {von Keyserlingk}}, \bibinfo {author} {\bibfnamefont {N.~Y.}\ \bibnamefont {Yao}}, \bibinfo {author} {\bibfnamefont {E.}~\bibnamefont {Demler}},\ and\ \bibinfo {author} {\bibfnamefont {M.~D.}\ \bibnamefont {Lukin}},\ }\bibfield  {title} {\bibinfo {title} {Observation of discrete time-crystalline order in a disordered dipolar many-body
			system},\ }\href {https://doi.org/10.1038/nature21426} {\bibfield  {journal} {\bibinfo  {journal} {Nature}\ }\textbf {\bibinfo {volume} {543}},\ \bibinfo {pages} {221} (\bibinfo {year} {2017})}\BibitemShut {NoStop}%
	\bibitem [{\citenamefont {Mi}\ \emph {et~al.}(2022)\citenamefont {Mi}, \citenamefont {Ippoliti}, \citenamefont {Quintana}, \citenamefont {Greene}, \citenamefont {Chen}, \citenamefont {Gross}, \citenamefont {Arute}, \citenamefont {Arya}, \citenamefont {Atalaya}, \citenamefont {Babbush} \emph {et~al.}}]{Mi2022}%
	\BibitemOpen
	\bibfield  {author} {\bibinfo {author} {\bibfnamefont {X.}~\bibnamefont {Mi}}, \bibinfo {author} {\bibfnamefont {M.}~\bibnamefont {Ippoliti}}, \bibinfo {author} {\bibfnamefont {C.}~\bibnamefont {Quintana}}, \bibinfo {author} {\bibfnamefont {A.}~\bibnamefont {Greene}}, \bibinfo {author} {\bibfnamefont {Z.}~\bibnamefont {Chen}}, \bibinfo {author} {\bibfnamefont {J.}~\bibnamefont {Gross}}, \bibinfo {author} {\bibfnamefont {F.}~\bibnamefont {Arute}}, \bibinfo {author} {\bibfnamefont {K.}~\bibnamefont {Arya}}, \bibinfo {author} {\bibfnamefont {J.}~\bibnamefont {Atalaya}}, \bibinfo {author} {\bibfnamefont {R.}~\bibnamefont {Babbush}}, \emph {et~al.},\ }\bibfield  {title} {\bibinfo {title} {Time-crystalline eigenstate order on a quantum processor},\ }\href {https://doi.org/10.1038/s41586-021-04257-w} {\bibfield  {journal} {\bibinfo  {journal} {Nature}\ }\textbf {\bibinfo {volume} {601}},\ \bibinfo {pages} {531} (\bibinfo {year} {2022})}\BibitemShut {NoStop}%
	\bibitem [{\citenamefont {King}\ \emph {et~al.}(2022)\citenamefont {King}, \citenamefont {Suzuki}, \citenamefont {Raymond}, \citenamefont {Zucca}, \citenamefont {Lanting}, \citenamefont {Altomare}, \citenamefont {Berkley}, \citenamefont {Ejtemaee}, \citenamefont {Hoskinson}, \citenamefont {Huang} \emph {et~al.}}]{King2022}%
	\BibitemOpen
	\bibfield  {author} {\bibinfo {author} {\bibfnamefont {A.~D.}\ \bibnamefont {King}}, \bibinfo {author} {\bibfnamefont {S.}~\bibnamefont {Suzuki}}, \bibinfo {author} {\bibfnamefont {J.}~\bibnamefont {Raymond}}, \bibinfo {author} {\bibfnamefont {A.}~\bibnamefont {Zucca}}, \bibinfo {author} {\bibfnamefont {T.}~\bibnamefont {Lanting}}, \bibinfo {author} {\bibfnamefont {F.}~\bibnamefont {Altomare}}, \bibinfo {author} {\bibfnamefont {A.~J.}\ \bibnamefont {Berkley}}, \bibinfo {author} {\bibfnamefont {S.}~\bibnamefont {Ejtemaee}}, \bibinfo {author} {\bibfnamefont {E.}~\bibnamefont {Hoskinson}}, \bibinfo {author} {\bibfnamefont {S.}~\bibnamefont {Huang}}, \emph {et~al.},\ }\bibfield  {title} {\bibinfo {title} {Coherent quantum annealing in a programmable 2,000{\thinspace}qubit ising chain},\ }\href {https://doi.org/10.1038/s41567-022-01741-6} {\bibfield  {journal} {\bibinfo  {journal} {Nature Physics}\ }\textbf {\bibinfo {volume} {18}},\ \bibinfo {pages} {1324} (\bibinfo {year} {2022})}\BibitemShut {NoStop}%
	\bibitem [{\citenamefont {Kim}\ \emph {et~al.}(2023)\citenamefont {Kim}, \citenamefont {Eddins}, \citenamefont {Anand}, \citenamefont {Wei}, \citenamefont {van~den Berg}, \citenamefont {Rosenblatt}, \citenamefont {Nayfeh}, \citenamefont {Wu}, \citenamefont {Zaletel}, \citenamefont {Temme},\ and\ \citenamefont {Kandala}}]{Kim2023}%
	\BibitemOpen
	\bibfield  {author} {\bibinfo {author} {\bibfnamefont {Y.}~\bibnamefont {Kim}}, \bibinfo {author} {\bibfnamefont {A.}~\bibnamefont {Eddins}}, \bibinfo {author} {\bibfnamefont {S.}~\bibnamefont {Anand}}, \bibinfo {author} {\bibfnamefont {K.~X.}\ \bibnamefont {Wei}}, \bibinfo {author} {\bibfnamefont {E.}~\bibnamefont {van~den Berg}}, \bibinfo {author} {\bibfnamefont {S.}~\bibnamefont {Rosenblatt}}, \bibinfo {author} {\bibfnamefont {H.}~\bibnamefont {Nayfeh}}, \bibinfo {author} {\bibfnamefont {Y.}~\bibnamefont {Wu}}, \bibinfo {author} {\bibfnamefont {M.}~\bibnamefont {Zaletel}}, \bibinfo {author} {\bibfnamefont {K.}~\bibnamefont {Temme}},\ and\ \bibinfo {author} {\bibfnamefont {A.}~\bibnamefont {Kandala}},\ }\bibfield  {title} {\bibinfo {title} {Evidence for the utility of quantum computing before fault tolerance},\ }\href {https://doi.org/10.1038/s41586-023-06096-3} {\bibfield  {journal} {\bibinfo  {journal} {Nature}\ }\textbf {\bibinfo {volume} {618}},\ \bibinfo {pages} {500} (\bibinfo {year}
		{2023})}\BibitemShut {NoStop}%
	\bibitem [{\citenamefont {Mi}\ \emph {et~al.}(2024)\citenamefont {Mi}, \citenamefont {Michailidis}, \citenamefont {Shabani}, \citenamefont {Miao}, \citenamefont {Klimov}, \citenamefont {Lloyd}, \citenamefont {Rosenberg}, \citenamefont {Acharya}, \citenamefont {Aleiner}, \citenamefont {Andersen} \emph {et~al.}}]{MiGoogle2024}%
	\BibitemOpen
	\bibfield  {author} {\bibinfo {author} {\bibfnamefont {X.}~\bibnamefont {Mi}}, \bibinfo {author} {\bibfnamefont {A.~A.}\ \bibnamefont {Michailidis}}, \bibinfo {author} {\bibfnamefont {S.}~\bibnamefont {Shabani}}, \bibinfo {author} {\bibfnamefont {K.~C.}\ \bibnamefont {Miao}}, \bibinfo {author} {\bibfnamefont {P.~V.}\ \bibnamefont {Klimov}}, \bibinfo {author} {\bibfnamefont {J.}~\bibnamefont {Lloyd}}, \bibinfo {author} {\bibfnamefont {E.}~\bibnamefont {Rosenberg}}, \bibinfo {author} {\bibfnamefont {R.}~\bibnamefont {Acharya}}, \bibinfo {author} {\bibfnamefont {I.}~\bibnamefont {Aleiner}}, \bibinfo {author} {\bibfnamefont {T.~I.}\ \bibnamefont {Andersen}}, \emph {et~al.},\ }\bibfield  {title} {\bibinfo {title} {Stable quantum-correlated many-body states through engineered dissipation},\ }\href {https://doi.org/10.1126/science.adh9932} {\bibfield  {journal} {\bibinfo  {journal} {Science}\ }\textbf {\bibinfo {volume} {383}},\ \bibinfo {pages} {1332} (\bibinfo {year} {2024})},\ \Eprint
	{https://arxiv.org/abs/https://www.science.org/doi/pdf/10.1126/science.adh9932} {https://www.science.org/doi/pdf/10.1126/science.adh9932} \BibitemShut {NoStop}%
	\bibitem [{\citenamefont {Baez}\ \emph {et~al.}(2020)\citenamefont {Baez}, \citenamefont {Goihl}, \citenamefont {Haferkamp}, \citenamefont {Bermejo-Vega}, \citenamefont {Gluza},\ and\ \citenamefont {Eisert}}]{Baez2020}%
	\BibitemOpen
	\bibfield  {author} {\bibinfo {author} {\bibfnamefont {M.~L.}\ \bibnamefont {Baez}}, \bibinfo {author} {\bibfnamefont {M.}~\bibnamefont {Goihl}}, \bibinfo {author} {\bibfnamefont {J.}~\bibnamefont {Haferkamp}}, \bibinfo {author} {\bibfnamefont {J.}~\bibnamefont {Bermejo-Vega}}, \bibinfo {author} {\bibfnamefont {M.}~\bibnamefont {Gluza}},\ and\ \bibinfo {author} {\bibfnamefont {J.}~\bibnamefont {Eisert}},\ }\bibfield  {title} {\bibinfo {title} {Dynamical structure factors of dynamical quantum simulators},\ }\href {https://doi.org/10.1073/pnas.2006103117} {\bibfield  {journal} {\bibinfo  {journal} {Proceedings of the National Academy of Sciences}\ }\textbf {\bibinfo {volume} {117}},\ \bibinfo {pages} {26123} (\bibinfo {year} {2020})}\BibitemShut {NoStop}%
	\bibitem [{\citenamefont {Will}\ \emph {et~al.}(2025)\citenamefont {Will}, \citenamefont {Cochran}, \citenamefont {Rosenberg}, \citenamefont {Jobst}, \citenamefont {Eassa}, \citenamefont {Roushan}, \citenamefont {Knap}, \citenamefont {Gammon-Smith},\ and\ \citenamefont {Pollmann}}]{Will2025}%
	\BibitemOpen
	\bibfield  {author} {\bibinfo {author} {\bibfnamefont {M.}~\bibnamefont {Will}}, \bibinfo {author} {\bibfnamefont {T.~A.}\ \bibnamefont {Cochran}}, \bibinfo {author} {\bibfnamefont {E.}~\bibnamefont {Rosenberg}}, \bibinfo {author} {\bibfnamefont {B.}~\bibnamefont {Jobst}}, \bibinfo {author} {\bibfnamefont {N.~M.}\ \bibnamefont {Eassa}}, \bibinfo {author} {\bibfnamefont {P.}~\bibnamefont {Roushan}}, \bibinfo {author} {\bibfnamefont {M.}~\bibnamefont {Knap}}, \bibinfo {author} {\bibfnamefont {A.}~\bibnamefont {Gammon-Smith}},\ and\ \bibinfo {author} {\bibfnamefont {F.}~\bibnamefont {Pollmann}},\ }\bibfield  {title} {\bibinfo {title} {Probing non-equilibrium topological order on a quantum processor},\ }\href {https://doi.org/10.1038/s41586-025-09456-3} {\bibfield  {journal} {\bibinfo  {journal} {Nature}\ }\textbf {\bibinfo {volume} {645}},\ \bibinfo {pages} {348} (\bibinfo {year} {2025})}\BibitemShut {NoStop}%
	\bibitem [{\citenamefont {Lee}\ \emph {et~al.}(2026)\citenamefont {Lee}, \citenamefont {Kumaran}, \citenamefont {Pokharel}, \citenamefont {Scheie}, \citenamefont {Sarkis}, \citenamefont {Tennant}, \citenamefont {Humble}, \citenamefont {Schleife}, \citenamefont {Kandala},\ and\ \citenamefont {Banerjee}}]{Lee2026}%
	\BibitemOpen
	\bibfield  {author} {\bibinfo {author} {\bibfnamefont {Y.-T.}\ \bibnamefont {Lee}}, \bibinfo {author} {\bibfnamefont {K.}~\bibnamefont {Kumaran}}, \bibinfo {author} {\bibfnamefont {B.}~\bibnamefont {Pokharel}}, \bibinfo {author} {\bibfnamefont {A.}~\bibnamefont {Scheie}}, \bibinfo {author} {\bibfnamefont {C.~L.}\ \bibnamefont {Sarkis}}, \bibinfo {author} {\bibfnamefont {D.~A.}\ \bibnamefont {Tennant}}, \bibinfo {author} {\bibfnamefont {T.}~\bibnamefont {Humble}}, \bibinfo {author} {\bibfnamefont {A.}~\bibnamefont {Schleife}}, \bibinfo {author} {\bibfnamefont {A.}~\bibnamefont {Kandala}},\ and\ \bibinfo {author} {\bibfnamefont {A.}~\bibnamefont {Banerjee}},\ }\href {https://arxiv.org/abs/2603.15608} {\bibinfo {title} {Benchmarking quantum simulation with neutron-scattering experiments}} (\bibinfo {year} {2026}),\ \Eprint {https://arxiv.org/abs/2603.15608} {arXiv:2603.15608 [quant-ph]} \BibitemShut {NoStop}%
	\bibitem [{\citenamefont {Zaliznyak}\ and\ \citenamefont {Tranquada}(2015)}]{Zaliznyak2015}%
	\BibitemOpen
	\bibfield  {author} {\bibinfo {author} {\bibfnamefont {I.~A.}\ \bibnamefont {Zaliznyak}}\ and\ \bibinfo {author} {\bibfnamefont {J.~M.}\ \bibnamefont {Tranquada}},\ }\bibinfo {title} {Neutron scattering and its application to strongly correlated systems},\ in\ \href {https://doi.org/10.1007/978-3-662-44133-6_7} {\emph {\bibinfo {booktitle} {Strongly Correlated Systems: Experimental Techniques}}},\ \bibinfo {editor} {edited by\ \bibinfo {editor} {\bibfnamefont {A.}~\bibnamefont {Avella}}\ and\ \bibinfo {editor} {\bibfnamefont {F.}~\bibnamefont {Mancini}}}\ (\bibinfo  {publisher} {Springer Berlin Heidelberg},\ \bibinfo {address} {Berlin, Heidelberg},\ \bibinfo {year} {2015})\ pp.\ \bibinfo {pages} {205--235}\BibitemShut {NoStop}%
	\bibitem [{\citenamefont {Devereaux}\ and\ \citenamefont {Hackl}(2007)}]{DevereauxandHackl2007}%
	\BibitemOpen
	\bibfield  {author} {\bibinfo {author} {\bibfnamefont {T.~P.}\ \bibnamefont {Devereaux}}\ and\ \bibinfo {author} {\bibfnamefont {R.}~\bibnamefont {Hackl}},\ }\bibfield  {title} {\bibinfo {title} {Inelastic light scattering from correlated electrons},\ }\href {https://doi.org/10.1103/RevModPhys.79.175} {\bibfield  {journal} {\bibinfo  {journal} {Rev. Mod. Phys.}\ }\textbf {\bibinfo {volume} {79}},\ \bibinfo {pages} {175} (\bibinfo {year} {2007})}\BibitemShut {NoStop}%
	\bibitem [{\citenamefont {F{\"o}lling}\ \emph {et~al.}(2005)\citenamefont {F{\"o}lling}, \citenamefont {Gerbier}, \citenamefont {Widera}, \citenamefont {Mandel}, \citenamefont {Gericke},\ and\ \citenamefont {Bloch}}]{Folling2005}%
	\BibitemOpen
	\bibfield  {author} {\bibinfo {author} {\bibfnamefont {S.}~\bibnamefont {F{\"o}lling}}, \bibinfo {author} {\bibfnamefont {F.}~\bibnamefont {Gerbier}}, \bibinfo {author} {\bibfnamefont {A.}~\bibnamefont {Widera}}, \bibinfo {author} {\bibfnamefont {O.}~\bibnamefont {Mandel}}, \bibinfo {author} {\bibfnamefont {T.}~\bibnamefont {Gericke}},\ and\ \bibinfo {author} {\bibfnamefont {I.}~\bibnamefont {Bloch}},\ }\bibfield  {title} {\bibinfo {title} {Spatial quantum noise interferometry in expanding ultracold atom clouds},\ }\href {https://doi.org/10.1038/nature03500} {\bibfield  {journal} {\bibinfo  {journal} {Nature}\ }\textbf {\bibinfo {volume} {434}},\ \bibinfo {pages} {481} (\bibinfo {year} {2005})}\BibitemShut {NoStop}%
	\bibitem [{\citenamefont {Reulet}\ \emph {et~al.}(2003)\citenamefont {Reulet}, \citenamefont {Senzier},\ and\ \citenamefont {Prober}}]{Reulet2003}%
	\BibitemOpen
	\bibfield  {author} {\bibinfo {author} {\bibfnamefont {B.}~\bibnamefont {Reulet}}, \bibinfo {author} {\bibfnamefont {J.}~\bibnamefont {Senzier}},\ and\ \bibinfo {author} {\bibfnamefont {D.~E.}\ \bibnamefont {Prober}},\ }\bibfield  {title} {\bibinfo {title} {Environmental effects in the third moment of voltage fluctuations in a tunnel junction},\ }\href {https://doi.org/10.1103/PhysRevLett.91.196601} {\bibfield  {journal} {\bibinfo  {journal} {Phys. Rev. Lett.}\ }\textbf {\bibinfo {volume} {91}},\ \bibinfo {pages} {196601} (\bibinfo {year} {2003})}\BibitemShut {NoStop}%
	\bibitem [{\citenamefont {Altman}\ \emph {et~al.}(2004)\citenamefont {Altman}, \citenamefont {Demler},\ and\ \citenamefont {Lukin}}]{Altman2004}%
	\BibitemOpen
	\bibfield  {author} {\bibinfo {author} {\bibfnamefont {E.}~\bibnamefont {Altman}}, \bibinfo {author} {\bibfnamefont {E.}~\bibnamefont {Demler}},\ and\ \bibinfo {author} {\bibfnamefont {M.~D.}\ \bibnamefont {Lukin}},\ }\bibfield  {title} {\bibinfo {title} {Probing many-body states of ultracold atoms via noise correlations},\ }\href {https://doi.org/10.1103/PhysRevA.70.013603} {\bibfield  {journal} {\bibinfo  {journal} {Phys. Rev. A}\ }\textbf {\bibinfo {volume} {70}},\ \bibinfo {pages} {013603} (\bibinfo {year} {2004})}\BibitemShut {NoStop}%
	\bibitem [{\citenamefont {Bakr}\ \emph {et~al.}(2009)\citenamefont {Bakr}, \citenamefont {Gillen}, \citenamefont {Peng}, \citenamefont {F{\"o}lling},\ and\ \citenamefont {Greiner}}]{Bakr2009}%
	\BibitemOpen
	\bibfield  {author} {\bibinfo {author} {\bibfnamefont {W.~S.}\ \bibnamefont {Bakr}}, \bibinfo {author} {\bibfnamefont {J.~I.}\ \bibnamefont {Gillen}}, \bibinfo {author} {\bibfnamefont {A.}~\bibnamefont {Peng}}, \bibinfo {author} {\bibfnamefont {S.}~\bibnamefont {F{\"o}lling}},\ and\ \bibinfo {author} {\bibfnamefont {M.}~\bibnamefont {Greiner}},\ }\bibfield  {title} {\bibinfo {title} {A quantum gas microscope for detecting single atoms in a hubbard-regime optical lattice},\ }\href {https://doi.org/10.1038/nature08482} {\bibfield  {journal} {\bibinfo  {journal} {Nature}\ }\textbf {\bibinfo {volume} {462}},\ \bibinfo {pages} {74} (\bibinfo {year} {2009})}\BibitemShut {NoStop}%
	\bibitem [{\citenamefont {Bando}\ \emph {et~al.}(2020)\citenamefont {Bando}, \citenamefont {Susa}, \citenamefont {Oshiyama}, \citenamefont {Shibata}, \citenamefont {Ohzeki}, \citenamefont {G\'omez-Ruiz}, \citenamefont {Lidar}, \citenamefont {Suzuki}, \citenamefont {del Campo},\ and\ \citenamefont {Nishimori}}]{Bando2020}%
	\BibitemOpen
	\bibfield  {author} {\bibinfo {author} {\bibfnamefont {Y.}~\bibnamefont {Bando}}, \bibinfo {author} {\bibfnamefont {Y.}~\bibnamefont {Susa}}, \bibinfo {author} {\bibfnamefont {H.}~\bibnamefont {Oshiyama}}, \bibinfo {author} {\bibfnamefont {N.}~\bibnamefont {Shibata}}, \bibinfo {author} {\bibfnamefont {M.}~\bibnamefont {Ohzeki}}, \bibinfo {author} {\bibfnamefont {F.~J.}\ \bibnamefont {G\'omez-Ruiz}}, \bibinfo {author} {\bibfnamefont {D.~A.}\ \bibnamefont {Lidar}}, \bibinfo {author} {\bibfnamefont {S.}~\bibnamefont {Suzuki}}, \bibinfo {author} {\bibfnamefont {A.}~\bibnamefont {del Campo}},\ and\ \bibinfo {author} {\bibfnamefont {H.}~\bibnamefont {Nishimori}},\ }\bibfield  {title} {\bibinfo {title} {Probing the universality of topological defect formation in a quantum annealer: Kibble-zurek mechanism and beyond},\ }\href {https://doi.org/10.1103/PhysRevResearch.2.033369} {\bibfield  {journal} {\bibinfo  {journal} {Phys. Rev. Res.}\ }\textbf {\bibinfo {volume} {2}},\ \bibinfo {pages} {033369} (\bibinfo {year}
		{2020})}\BibitemShut {NoStop}%
	\bibitem [{\citenamefont {Quan}\ \emph {et~al.}(2006)\citenamefont {Quan}, \citenamefont {Song}, \citenamefont {Liu}, \citenamefont {Zanardi},\ and\ \citenamefont {Sun}}]{Quan2006}%
	\BibitemOpen
	\bibfield  {author} {\bibinfo {author} {\bibfnamefont {H.~T.}\ \bibnamefont {Quan}}, \bibinfo {author} {\bibfnamefont {Z.}~\bibnamefont {Song}}, \bibinfo {author} {\bibfnamefont {X.~F.}\ \bibnamefont {Liu}}, \bibinfo {author} {\bibfnamefont {P.}~\bibnamefont {Zanardi}},\ and\ \bibinfo {author} {\bibfnamefont {C.~P.}\ \bibnamefont {Sun}},\ }\bibfield  {title} {\bibinfo {title} {Decay of loschmidt echo enhanced by quantum criticality},\ }\bibfield  {journal} {\bibinfo  {journal} {Physical Review Letters}\ }\textbf {\bibinfo {volume} {96}},\ \href {https://doi.org/10.1103/physrevlett.96.140604} {10.1103/physrevlett.96.140604} (\bibinfo {year} {2006})\BibitemShut {NoStop}%
	\bibitem [{\citenamefont {Cucchietti}\ \emph {et~al.}(2007)\citenamefont {Cucchietti}, \citenamefont {Fernandez-Vidal},\ and\ \citenamefont {Paz}}]{Cucchietti2007}%
	\BibitemOpen
	\bibfield  {author} {\bibinfo {author} {\bibfnamefont {F.~M.}\ \bibnamefont {Cucchietti}}, \bibinfo {author} {\bibfnamefont {S.}~\bibnamefont {Fernandez-Vidal}},\ and\ \bibinfo {author} {\bibfnamefont {J.~P.}\ \bibnamefont {Paz}},\ }\bibfield  {title} {\bibinfo {title} {Universal decoherence induced by an environmental quantum phase transition},\ }\href {https://doi.org/10.1103/PhysRevA.75.032337} {\bibfield  {journal} {\bibinfo  {journal} {Phys. Rev. A}\ }\textbf {\bibinfo {volume} {75}},\ \bibinfo {pages} {032337} (\bibinfo {year} {2007})}\BibitemShut {NoStop}%
	\bibitem [{\citenamefont {Haikka}\ \emph {et~al.}(2012)\citenamefont {Haikka}, \citenamefont {Goold}, \citenamefont {McEndoo}, \citenamefont {Plastina},\ and\ \citenamefont {Maniscalco}}]{Haikka2012}%
	\BibitemOpen
	\bibfield  {author} {\bibinfo {author} {\bibfnamefont {P.}~\bibnamefont {Haikka}}, \bibinfo {author} {\bibfnamefont {J.}~\bibnamefont {Goold}}, \bibinfo {author} {\bibfnamefont {S.}~\bibnamefont {McEndoo}}, \bibinfo {author} {\bibfnamefont {F.}~\bibnamefont {Plastina}},\ and\ \bibinfo {author} {\bibfnamefont {S.}~\bibnamefont {Maniscalco}},\ }\bibfield  {title} {\bibinfo {title} {Non-markovianity, loschmidt echo, and criticality: A unified picture},\ }\bibfield  {journal} {\bibinfo  {journal} {Physical Review A}\ }\textbf {\bibinfo {volume} {85}},\ \href {https://doi.org/10.1103/physreva.85.060101} {10.1103/physreva.85.060101} (\bibinfo {year} {2012})\BibitemShut {NoStop}%
	\bibitem [{\citenamefont {Damski}\ \emph {et~al.}(2011)\citenamefont {Damski}, \citenamefont {Quan},\ and\ \citenamefont {Zurek}}]{Damski2011}%
	\BibitemOpen
	\bibfield  {author} {\bibinfo {author} {\bibfnamefont {B.}~\bibnamefont {Damski}}, \bibinfo {author} {\bibfnamefont {H.~T.}\ \bibnamefont {Quan}},\ and\ \bibinfo {author} {\bibfnamefont {W.~H.}\ \bibnamefont {Zurek}},\ }\bibfield  {title} {\bibinfo {title} {Critical dynamics of decoherence},\ }\href {https://doi.org/10.1103/PhysRevA.83.062104} {\bibfield  {journal} {\bibinfo  {journal} {Phys. Rev. A}\ }\textbf {\bibinfo {volume} {83}},\ \bibinfo {pages} {062104} (\bibinfo {year} {2011})}\BibitemShut {NoStop}%
	\bibitem [{\citenamefont {Giorgi}\ \emph {et~al.}(2019)\citenamefont {Giorgi}, \citenamefont {Longhi}, \citenamefont {Cabot},\ and\ \citenamefont {Zambrini}}]{Giorgi2019}%
	\BibitemOpen
	\bibfield  {author} {\bibinfo {author} {\bibfnamefont {G.~L.}\ \bibnamefont {Giorgi}}, \bibinfo {author} {\bibfnamefont {S.}~\bibnamefont {Longhi}}, \bibinfo {author} {\bibfnamefont {A.}~\bibnamefont {Cabot}},\ and\ \bibinfo {author} {\bibfnamefont {R.}~\bibnamefont {Zambrini}},\ }\bibfield  {title} {\bibinfo {title} {Quantum probing topological phase transitions by non‐markovianity},\ }\bibfield  {journal} {\bibinfo  {journal} {Annalen der Physik}\ }\textbf {\bibinfo {volume} {531}},\ \href {https://doi.org/10.1002/andp.201900307} {10.1002/andp.201900307} (\bibinfo {year} {2019})\BibitemShut {NoStop}%
	\bibitem [{\citenamefont {Mirkin}\ \emph {et~al.}(2021)\citenamefont {Mirkin}, \citenamefont {Wisniacki}, \citenamefont {Villar},\ and\ \citenamefont {Lombardo}}]{Mirkin2021}%
	\BibitemOpen
	\bibfield  {author} {\bibinfo {author} {\bibfnamefont {N.}~\bibnamefont {Mirkin}}, \bibinfo {author} {\bibfnamefont {D.~A.}\ \bibnamefont {Wisniacki}}, \bibinfo {author} {\bibfnamefont {P.~I.}\ \bibnamefont {Villar}},\ and\ \bibinfo {author} {\bibfnamefont {F.~C.}\ \bibnamefont {Lombardo}},\ }\bibfield  {title} {\bibinfo {title} {Sensing quantum chaos through the non-unitary geometric phase},\ }\href {https://doi.org/10.1088/2058-9565/ac1e37} {\bibfield  {journal} {\bibinfo  {journal} {Quantum Sci. Technol.}\ }\textbf {\bibinfo {volume} {6}},\ \bibinfo {pages} {045018} (\bibinfo {year} {2021})}\BibitemShut {NoStop}%
	\bibitem [{\citenamefont {Roberts}\ \emph {et~al.}(2024)\citenamefont {Roberts}, \citenamefont {Vrajitoarea}, \citenamefont {Saxberg}, \citenamefont {Panetta}, \citenamefont {Simon},\ and\ \citenamefont {Schuster}}]{Roberts2024}%
	\BibitemOpen
	\bibfield  {author} {\bibinfo {author} {\bibfnamefont {G.}~\bibnamefont {Roberts}}, \bibinfo {author} {\bibfnamefont {A.}~\bibnamefont {Vrajitoarea}}, \bibinfo {author} {\bibfnamefont {B.}~\bibnamefont {Saxberg}}, \bibinfo {author} {\bibfnamefont {M.~G.}\ \bibnamefont {Panetta}}, \bibinfo {author} {\bibfnamefont {J.}~\bibnamefont {Simon}},\ and\ \bibinfo {author} {\bibfnamefont {D.~I.}\ \bibnamefont {Schuster}},\ }\bibfield  {title} {\bibinfo {title} {Manybody interferometry of quantum fluids},\ }\href {https://doi.org/10.1126/sciadv.ado1069} {\bibfield  {journal} {\bibinfo  {journal} {Science Advances}\ }\textbf {\bibinfo {volume} {10}},\ \bibinfo {pages} {eado1069} (\bibinfo {year} {2024})},\ \Eprint {https://arxiv.org/abs/https://www.science.org/doi/pdf/10.1126/sciadv.ado1069} {https://www.science.org/doi/pdf/10.1126/sciadv.ado1069} \BibitemShut {NoStop}%
	\bibitem [{\citenamefont {Keefe}\ \emph {et~al.}(2025)\citenamefont {Keefe}, \citenamefont {Agarwal},\ and\ \citenamefont {Kamal}}]{keefe2025}%
	\BibitemOpen
	\bibfield  {author} {\bibinfo {author} {\bibfnamefont {A.}~\bibnamefont {Keefe}}, \bibinfo {author} {\bibfnamefont {N.}~\bibnamefont {Agarwal}},\ and\ \bibinfo {author} {\bibfnamefont {A.}~\bibnamefont {Kamal}},\ }\bibfield  {title} {\bibinfo {title} {Quantifying spectral signatures of non-markovianity beyond the born-redfield master equation},\ }\href {https://doi.org/https://doi.org/10.22331/q-2025-09-24-1863} {\bibfield  {journal} {\bibinfo  {journal} {Quantum}\ }\textbf {\bibinfo {volume} {9}},\ \bibinfo {pages} {1863} (\bibinfo {year} {2025})}\BibitemShut {NoStop}%
	\bibitem [{\citenamefont {Jordan}\ and\ \citenamefont {Wigner}(1928)}]{Jordan_1928}%
	\BibitemOpen
	\bibfield  {author} {\bibinfo {author} {\bibfnamefont {P.}~\bibnamefont {Jordan}}\ and\ \bibinfo {author} {\bibfnamefont {E.}~\bibnamefont {Wigner}},\ }\bibfield  {title} {\bibinfo {title} {\"uber das paulische \"aquivalenzverbot},\ }\href@noop {} {\bibfield  {journal} {\bibinfo  {journal} {Zeitschrift f\"ur Physik}\ }\textbf {\bibinfo {volume} {47}} (\bibinfo {year} {1928})}\BibitemShut {NoStop}%
	\bibitem [{Note1()}]{Note1}%
	\BibitemOpen
	\bibinfo {note} {Beyond quadratic order, normal ordering contains additional terms to cancel all bubble diagrams \cite {Bowen:2025kyo}.}\BibitemShut {Stop}%
	\bibitem [{Note2()}]{Note2}%
	\BibitemOpen
	\bibinfo {note} {We normal order $ \protect \hat {\protect \mathcal {O}}_{\protect \text {E}} $ and move a $ \mathop {:} \protect \hat {\gamma }_{k}^{\dagger }\protect \hat {\gamma }_{k} \mathop {:} $ term that leads to spurious thermal bubbles in the second-order environment correlation function to the free Hamiltonian, where it becomes a system-dependent frequency correction to the TFIM spectrum that ultimately vanishes in the thermodynamic limit (see \protect \cref {app:diagTFIM}).}\BibitemShut {Stop}%
	\bibitem [{Note3()}]{Note3}%
	\BibitemOpen
	\bibinfo {note} {The absence of a unitary correction (Lamb shift) in the Redfield equation is a direct result of the QND interaction.}\BibitemShut {Stop}%
	\bibitem [{\citenamefont {de~Jong}\ and\ \citenamefont {Beenakker}(1997)}]{JongBeenakker}%
	\BibitemOpen
	\bibfield  {author} {\bibinfo {author} {\bibfnamefont {M.~J.~M.}\ \bibnamefont {de~Jong}}\ and\ \bibinfo {author} {\bibfnamefont {C.~W.~J.}\ \bibnamefont {Beenakker}},\ }\bibfield  {title} {\bibinfo {title} {Shot noise in mesoscopic systems},\ }in\ \href@noop {} {\emph {\bibinfo {booktitle} {Mesoscopic Electron Transport}}},\ \bibinfo {series} {NATO ASI Series}, Vol.\ \bibinfo {volume} {345},\ \bibinfo {editor} {edited by\ \bibinfo {editor} {\bibfnamefont {L.}~\bibnamefont {Sohn}}, \bibinfo {editor} {\bibfnamefont {L.}~\bibnamefont {Kouwenhoven}},\ and\ \bibinfo {editor} {\bibfnamefont {G.}~\bibnamefont {Schoen}}}\ (\bibinfo  {publisher} {Kluwer Academic Publishers},\ \bibinfo {address} {Dordrecht},\ \bibinfo {year} {1997})\ pp.\ \bibinfo {pages} {225--258}\BibitemShut {NoStop}%
	\bibitem [{Note4()}]{Note4}%
	\BibitemOpen
	\bibinfo {note} {In this case, the third singular value is negligible compared to the first two.}\BibitemShut {Stop}%
	\bibitem [{Note5()}]{Note5}%
	\BibitemOpen
	\bibinfo {note} {The overlapping vectors in $ \protect \mathcal {F}_{\protect \mathbf {u},2}(\mu , T) $ reflect the curvature of $ \protect \mathcal {D}(\mu , T) $ in the full three-dimensional fixed-point space, where the construction of the Jacobian does not commute with the projection onto the principal plane.}\BibitemShut {Stop}%
	\bibitem [{\citenamefont {Catuneanu}\ \emph {et~al.}(2019)\citenamefont {Catuneanu}, \citenamefont {S\o{}rensen},\ and\ \citenamefont {Kee}}]{Catuneanu2019}%
	\BibitemOpen
	\bibfield  {author} {\bibinfo {author} {\bibfnamefont {A.}~\bibnamefont {Catuneanu}}, \bibinfo {author} {\bibfnamefont {E.~S.}\ \bibnamefont {S\o{}rensen}},\ and\ \bibinfo {author} {\bibfnamefont {H.-Y.}\ \bibnamefont {Kee}},\ }\bibfield  {title} {\bibinfo {title} {Nonlocal string order parameter in the $s=\frac{1}{2}$ kitaev-heisenberg ladder},\ }\href {https://doi.org/10.1103/PhysRevB.99.195112} {\bibfield  {journal} {\bibinfo  {journal} {Phys. Rev. B}\ }\textbf {\bibinfo {volume} {99}},\ \bibinfo {pages} {195112} (\bibinfo {year} {2019})}\BibitemShut {NoStop}%
	\bibitem [{\citenamefont {Po}\ and\ \citenamefont {Zhou}(2015)}]{Po2015}%
	\BibitemOpen
	\bibfield  {author} {\bibinfo {author} {\bibfnamefont {H.~C.}\ \bibnamefont {Po}}\ and\ \bibinfo {author} {\bibfnamefont {Q.}~\bibnamefont {Zhou}},\ }\bibfield  {title} {\bibinfo {title} {A two-dimensional algebraic quantum liquid produced by an atomic simulator of the quantum lifshitz model},\ }\href {https://doi.org/10.1038/ncomms9012} {\bibfield  {journal} {\bibinfo  {journal} {Nature Communications}\ }\textbf {\bibinfo {volume} {6}},\ \bibinfo {pages} {8012} (\bibinfo {year} {2015})}\BibitemShut {NoStop}%
	\bibitem [{\citenamefont {Bowen}\ \emph {et~al.}(2026)\citenamefont {Bowen}, \citenamefont {Farah}, \citenamefont {Chaykov},\ and\ \citenamefont {Agarwal}}]{Bowen:2025kyo}%
	\BibitemOpen
	\bibfield  {author} {\bibinfo {author} {\bibfnamefont {B.}~\bibnamefont {Bowen}}, \bibinfo {author} {\bibfnamefont {A.}~\bibnamefont {Farah}}, \bibinfo {author} {\bibfnamefont {S.}~\bibnamefont {Chaykov}},\ and\ \bibinfo {author} {\bibfnamefont {N.}~\bibnamefont {Agarwal}},\ }\bibfield  {title} {\bibinfo {title} {{Mutual information as a measure of renormalizability}},\ }\href {https://doi.org/10.1007/JHEP06(2026)136} {\bibfield  {journal} {\bibinfo  {journal} {JHEP}\ }\textbf {\bibinfo {volume} {06}},\ \bibinfo {pages} {136}},\ \Eprint {https://arxiv.org/abs/2511.09625} {arXiv:2511.09625 [hep-th]} \BibitemShut {NoStop}%
	\bibitem [{\citenamefont {Sachdev}(2011)}]{Sachdev2011}%
	\BibitemOpen
	\bibfield  {author} {\bibinfo {author} {\bibfnamefont {S.}~\bibnamefont {Sachdev}},\ }\href {https://doi.org/10.1017/CBO9780511973765} {\emph {\bibinfo {title} {Quantum Phase Transitions}}},\ \bibinfo {edition} {2nd}\ ed.\ (\bibinfo  {publisher} {Cambridge University Press},\ \bibinfo {year} {2011})\BibitemShut {NoStop}%
	\bibitem [{\citenamefont {Bowen}\ \emph {et~al.}(2025)\citenamefont {Bowen}, \citenamefont {Agarwal},\ and\ \citenamefont {Kamal}}]{Bowen:2024emo}%
	\BibitemOpen
	\bibfield  {author} {\bibinfo {author} {\bibfnamefont {B.}~\bibnamefont {Bowen}}, \bibinfo {author} {\bibfnamefont {N.}~\bibnamefont {Agarwal}},\ and\ \bibinfo {author} {\bibfnamefont {A.}~\bibnamefont {Kamal}},\ }\bibfield  {title} {\bibinfo {title} {{Open system dynamics in interacting quantum field theories}},\ }\href {https://doi.org/10.1103/kjph-9z8l} {\bibfield  {journal} {\bibinfo  {journal} {Phys. Rev. Res.}\ }\textbf {\bibinfo {volume} {7}},\ \bibinfo {pages} {043311} (\bibinfo {year} {2025})},\ \Eprint {https://arxiv.org/abs/2403.18907} {arXiv:2403.18907 [hep-th]} \BibitemShut {NoStop}%
	\bibitem [{Note6()}]{Note6}%
	\BibitemOpen
	\bibinfo {note} {Had we directly used \protect \cref {eq:EnvironmentOperator} to compute the correlation function, we would simply have to use that, in the thermodynamic limit, $ L \delta _{mn}|_{n=m} \to 2 \pi \delta (k_{m}-k_{n})|_{n=m} = 2 \pi \delta (0) $, to obtain the same result.}\BibitemShut {Stop}%
	\bibitem [{\citenamefont {Itzykson}\ and\ \citenamefont {Drouffe}(1989)}]{Itzykson:1989sx}%
	\BibitemOpen
	\bibfield  {author} {\bibinfo {author} {\bibfnamefont {C.}~\bibnamefont {Itzykson}}\ and\ \bibinfo {author} {\bibfnamefont {J.~M.}\ \bibnamefont {Drouffe}},\ }\href {https://doi.org/10.1017/CBO9780511622779} {\emph {\bibinfo {title} {{Statistical Field Theory. Vol. 1: from Brownian motion to renormalization and lattice gauge theory}}}},\ Cambridge Monographs on Mathematical Physics\ (\bibinfo  {publisher} {CUP},\ \bibinfo {year} {1989})\BibitemShut {NoStop}%
	\bibitem [{\citenamefont {Di~Francesco}\ \emph {et~al.}(1997)\citenamefont {Di~Francesco}, \citenamefont {Mathieu},\ and\ \citenamefont {Senechal}}]{DiFrancesco:1997nk}%
	\BibitemOpen
	\bibfield  {author} {\bibinfo {author} {\bibfnamefont {P.}~\bibnamefont {Di~Francesco}}, \bibinfo {author} {\bibfnamefont {P.}~\bibnamefont {Mathieu}},\ and\ \bibinfo {author} {\bibfnamefont {D.}~\bibnamefont {Senechal}},\ }\href {https://doi.org/10.1007/978-1-4612-2256-9} {\emph {\bibinfo {title} {{Conformal Field Theory}}}},\ Graduate Texts in Contemporary Physics\ (\bibinfo  {publisher} {Springer-Verlag},\ \bibinfo {address} {New York},\ \bibinfo {year} {1997})\BibitemShut {NoStop}%
	\bibitem [{Note7()}]{Note7}%
	\BibitemOpen
	\bibinfo {note} {This ensures that a positively oriented contour picks up the pole.}\BibitemShut {Stop}%
\end{thebibliography}
\end{document}